\documentclass[12pt, draftclsnofoot, onecolumn]{IEEEtran}

\usepackage{cite}
\usepackage{amsmath,amssymb,amsfonts,bm}
\usepackage{algorithm}
\usepackage{algorithmic}
\usepackage{graphicx}
\usepackage{epstopdf}
\usepackage{booktabs}
\usepackage{hhline}
\usepackage{makecell}
\usepackage{multirow}
\usepackage{textcomp}
\usepackage{xcolor}
\usepackage[colorlinks  = true,
            linkcolor   = black,
            urlcolor    = magenta,
            anchorcolor = blue,
            bookmarks   = false
            ]{hyperref}

\def\BibTeX{{\rm B\kern-.05em{\sc i\kern-.025em b}\kern-.08em
    T\kern-.1667em\lower.7ex\hbox{E}\kern-.125emX}}

\DeclareMathOperator*{\ReLU}{ReLU}
\DeclareMathOperator*{\LReLU}{LReLU}
\DeclareMathOperator*{\PReLU}{PReLU}

\setcellgapes{2pt}

\begin{document}

\title{Binarized Aggregated Network with Quantization: Flexible Deep Learning Deployment for CSI Feedback in Massive MIMO System}

\author{

Zhilin Lu, Xudong Zhang, Hongyi He, Jintao Wang,~\IEEEmembership{Senior Member,~IEEE} and Jian Song,~\IEEEmembership{Fellow,~IEEE}

\thanks{
This work was supported in part by the National Key R\&D Program of China under Grant 2017YFE0112300.

The authors are with the Department of Electronic Engineering, Tsinghua University, and Beijing National Research Center for Information Science and Technology (BNRist), Beijing 100084, China. (e-mail: \{luzl18, zhang-xd18, hehy18\}@mails.tsinghua.edu.cn; \{wangjintao, jsong\}@tsinghua.edu.cn). Corresponding author: Jintao Wang.

The key results of this paper can be reproduced with github repository \href{https://github.com/Kylin9511/ACRNet}{https://github.com/Kylin9511/ACRNet}.
}% <-this % stops a space
% \thanks{Manuscript received XXX, XX, 2015; revised XXX, XX, 2015.}
}

%\markboth{IEEE Transactions on Vehicular Technology,~Vol.~XX, No.~XX, XXX~2020}
{}
%{Shell \MakeLowercase{\textit{et al.}}: Bare Demo of IEEEtran.cls for Journals}

\maketitle

\begin{abstract}
Massive multiple-input multiple-output (MIMO) is one of the key techniques to achieve better spectrum and energy efficiency in 5G system. The channel state information (CSI) needs to be fed back from the user equipment to the base station in frequency division duplexing (FDD) mode. However, the overhead of the direct feedback is unacceptable due to the large antenna array in massive MIMO system. Recently, deep learning is widely adopted to the compressed CSI feedback task and proved to be effective. In this paper, a novel network named aggregated channel reconstruction network (ACRNet) is designed to boost the feedback performance with network aggregation and parametric rectified linear unit (PReLU) activation. The practical deployment of the feedback network in the communication system is also considered. Specifically, the elastic feedback scheme is proposed to flexibly adapt the network to meet different resource limitations. Besides, the network binarization technique is combined with the feature quantization for lightweight and practical deployment. Experiments show that the proposed ACRNet outperforms loads of previous state-of-the-art networks, providing a neat feedback solution with high performance, low cost and impressive flexibility.
\end{abstract}

\newpage

\IEEEpeerreviewmaketitle

\section{Introduction} \label{Section1}

\IEEEPARstart{M}{assive} multiple-input multiple-output (MIMO) is widely regarded as one of the key techniques in the fifth-generation wireless communication system \cite{boccardi2014five}. With larger antenna array, massive MIMO is able to boost both spectrum and energy efficiency \cite{lu2014overview}. The downlink channel state information (CSI) needs to be obtained at the base station (BS) so that the MIMO system can acquire the performance gain with beamforming \cite{molisch2017hybrid}. In frequency division duplexing (FDD) mode, downlink CSI is usually estimated at the user equipment (UE) and fed back to the BS due to the channel non-reciprocity. However, the dimension of CSI matrix is sharply increased in massive MIMO system. The bandwidth consumed by CSI feedback is therefore unacceptable.

To address the above challenge, the CSI matrix should be compressed before feedback to reduce the overhead. The traditional compressed sensing (CS) method can not work well since the channel matrix is not sparse enough under large compression ratio \cite{wen2018deep}. In addition, the measurement matrix is usually non-optimal and the recovery is time consuming. On the other hand, deep learning (DL) has achieved great success in computer vision and signal processing \cite{lecun2015deep, voulodimos2018deep, young2018recent}. Specifically, DL based methods have dominated the image compression task \cite{toderici2017full, li2018learning}. This motivates the researchers to compress the CSI matrix with the neural network (NN) as well.

The CsiNet introduced in \cite{wen2018deep} is the first DL based CSI feedback algorithm. Basically, the CSI matrix is compressed by an auto-encoder at the UE to reduce the feedback overhead. After that, the compressed CSI feature is fed back to the BS, where the original CSI is recovered from it by an auto-decoder. Experiments show that the recovery of the CSI matrix could be rather precise. With the tremendous superiority against traditional CS based methods, CsiNet becomes the new state-of-the-art algorithm for massive MIMO CSI feedback.

On the basis of CsiNet, later researches on massive MIMO CSI feedback mainly focus on two fields, including the extension of the system model and the optimization of the feedback network.

On the one hand, many works are devoted to solve the feedback problem on different system model assumptions, which bring extra limitation or information to the fundamental feedback scenario. Time varying CSI feedback is considered in \cite{wang2018deep} using the recurrent neural network (RNN). \cite{li2020spatio} and \cite{liu2020markovian} further explore the time varying feedback with the spatial-temporal representation and the Markovian model driven network. Besides, correlation between uplink and downlink CSI is utilized in \cite{yang2019deep} and \cite{liu2019exploiting}. Multi-user cooperative feedback is considered in \cite{mashhadi2020distributed} and \cite{guo2020dl}. Moreover, denoise modules are introduced in \cite{ye2020deep} and \cite{sun2020ancinet} to deal with non-linear effect of imperfect feedback. Network safety against adversarial attack is improved in \cite{liu2020adversarial}. In addition, some joint optimized models are designed to combine CSI feedback with channel estimation \cite{guo2021canet, mashhadi2020deep, chen2020deep} or beamforming \cite{guo2020deep,elbir2020deep,sohrabi2021deep}. These models optimize the feedback block together with its upstream or downstream communication blocks to achieve better integral system performance.

On the other hand, a series of works focus on the optimization of the feedback network in order to improve the feedback performance or the deployment practicality. Note that the advanced algorithms proposed in these works can be inspiring for most of the aforementioned scenarios. The reason is that better performance and easier deployment are the common pursuit for any feedback system models.

There are many enlightening works towards better feedback performance. ReNet proposed in \cite{liang2020deep} combines NN decoder with CS based encoder. Multi-resolution aided CRNet achieves better feedback performance with the advanced training scheme in \cite{lu2020multi}. CsiNetPlus introduced in \cite{guo2020convolutional} reduces the feedback error with architecture transformation and larger convolution kernel. Non-local block is proved to be effective for feedback task in DS-NLCsiNet \cite{yu2020ds}. Squeeze and excitation (SE) block is added to the original CsiNet for better CSI feature extraction in \cite{cai2019attention}. Variational auto encoder (VAE) aided PRVNet is proposed in \cite{hussien2020prvnet} and achieves impressive CSI recovery performance.

At the mean while, it is actually quite challenging to deploy the feedback network efficiently in the real-world communication system. Instead of simply aiming at higher feedback accuracy, some existing works try to optimize the feedback network towards more practical deployment. Lightweight feedback network design is necessary due to the limited resources at both the UE and BS. Network sparse pruning is introduced to the feedback task in \cite{guo2020compression}. The depth-wise convolution and the channel shuffle is used to reduce the network complexity in \cite{cao2020lightweight}. An extremely light weight binary neural network (BNN) named BCsiNet is proposed in \cite{lu2021binary}, achieving over 30$\times$ memory saving and 2$\times$ acceleration compared with the original CsiNet \cite{wen2018deep}. Additionally, the feature quantization is indispensable since the uplink feedback is based on the bit stream instead of the float number. Uniform quantization is applied in JCNet \cite{lu2019bit} and the quantization with entropy encoder is proposed in DeepCMC \cite{yang2019deepcmc}. $\mu$-law quantization with learnable offset is introduced in \cite{guo2020convolutional} to reduce the information loss of the quantization.

However, previous research misses out on the diversity of the CSI spatial patterns. By designing advanced feedback network that can extract various spatial patterns better, the feedback performance can be further optimized. More importantly, the flexible and practical deployment of the feedback network needs further research as well. The existing feedback networks can not easily adapt to different resource limitations since their network complexities are usually fixed. The compatibility of deployment techniques like BNN and quantization is also unknown.

As a matter of fact, it is a common academic paradigm to design the elastic network architecture with expandable complexity. Take ResNet series \cite{he2016deep} as an example, lightweight ResNet18-half is suitable for real time inference on portable device while heavy ResNet101 is a better choice if the inference is offline on the GPU server. However, the elastic network design customized for the feedback task is still missing. As a result, users might get confused about how to adjust the network during the deployment if the hardware resource at the UE or BS is deficient or excessive.

\begin{figure}[t]
\centering
\includegraphics[width=\textwidth]{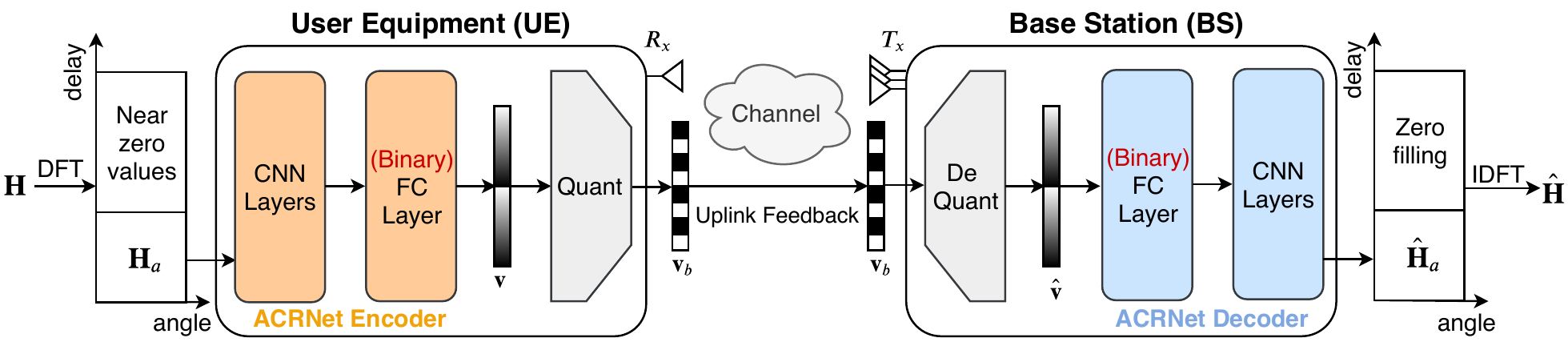}
\caption{Overview of downlink CSI feedback pipeline. The CSI matrix is first compressed into a feature vector $\mathbf{v}$ by the ACRNet encoder at the UE. Then the bitstream $\mathbf{v}_b$ is generated from $\mathbf{v}$ by a quantizer for uplink feedback. After that, the feature vector $\hat{\mathbf{v}}$ is dequantized from the received $\mathbf{v}_b$ and the original CSI matrix is reconstructed by the ACRNet decoder at the BS.}
\label{System Model}
\end{figure}

In this paper, we design a novel neural network named aggregated channel reconstruction network (ACRNet) to solve the challenges above. The pipeline of the ACRNet aided CSI feedback is given in Fig. \ref{System Model}. Both performance optimization and deployment optimization are considered for CSI feedback. The proposed ACRNet outperforms series of previous state-of-the-art feedback networks. Practical network expansion scheme is studied and binarization with quantization is proved to be effective. The main contributions of this paper are listed below.
\begin{itemize}
  \item In order to improve the feedback performance, the proposed ACRNet takes the advantage of network aggregation to learn various CSI patterns better. Additionally, learnable activation function is applied so that each convolutional group in ACRNet processes independent activation slope. This further enriches the CSI feature diversity brought by the network aggregation technique.
  \item In order to deploy the feedback network practically, the effective network expansion scheme is designed based on the characteristic of CSI matrix. By expanding the aggregated groups in ACRNet, better performance can be easily exchanged with reasonable extra complexity. To the best of our knowledge, we are the first to study the elastic network architecture for CSI feedback.
  \item Binarization technique is first combined with vector quantization to further improve the deployment flexibility. The binarized quantization scheme can work together with the elastic ACRNet design, providing flexible feedback deployment strategy for the communication system with high performance and low cost.
\end{itemize}

The rest of the paper is arranged as follows. Section \ref{Section2} introduces the system model of CSI feedback. Section \ref{Section3} explains how the design of ACRNet improves the feedback performance in details. Section \ref{Section4} presents the flexible deep learning deployment strategies for CSI feedback, including the network expansion and binarization with quantization. The numerical results and analysis are given in section \ref{Section5} and the conclusion is drawn in section \ref{Section6}.

\section{System Model} \label{Section2}

In this paper, a single cell massive MIMO system is adopted, where the BS equips $N_t$ transmitting antennas and the UE equips $N_r$ receiving antennas. Note that $N_t \gg N_r$ and $N_r$ is set to $1$ for simplicity. An orthogonal frequency division multiplexing (OFDM) system with $N_c$ sub-carriers is considered. The received signal $y_n$ on the $n^\text{th}$ sub-carrier ($n \in \{1, 2, \cdots, N_c\}$) at the UE can be expressed as follows:

\begin{equation}
    y_n = \textbf{h}_n^H\mathbf{p}_nx_n+z_n,
\end{equation}
where $x_n \in \mathbb{C}$ and $z_n \in \mathbb{C}$ are the transmitted symbol and additive Gaussian noise on the $n^\text{th}$ sub-carrier. $\mathbf{h}_n \in \mathbb{C}^{N_t\times 1}$ is the downlink channel response vector and $\mathbf{p}_n \in \mathbb{C}^{N_t\times 1}$ is the corresponding precoding vector. Since the uplink and downlink channel is not symmetric, $\mathbf{h}_n$ has to be estimated at the UE and fed back to the BS. We can concatenate all the channel response vectors into an overall downlink channel matrix $\mathbf{H}=[\mathbf{h}_1, \mathbf{h}_2, \cdots, \mathbf{h}_{N_c}]^H$. It is obvious that $\mathbf{H}$ contains $2\times N_c \times N_t$ float numbers, which is unacceptably large for direct feedback in massive MIMO system.

In order to reduce the size of the CSI matrix, we transfer $\mathbf{H}$ from the spatial-frequency domain to the angular-delay domain with discrete Fourier transform (DFT) as \cite{lu2020multi}.

\begin{equation} \label{eq2}
    \mathbf{H}' = \mathbf{F}_c\mathbf{H}\mathbf{F}_t^H,
\end{equation}
where $\mathbf{F}_c \in \mathbb{C}^{N_c\times N_c}$ and $\mathbf{F}_t \in \mathbb{C}^{N_t \times N_t}$ are the DFT transform matrices, respectively. $\mathbf{H}'$ is the angular-delay domain CSI matrix generated by DFT. As depicted in Fig. \ref{System Model}, the majority of the elements in $\mathbf{H}'$ are zero or near-zero, which can be omitted during feedback. More precisely, only the first $N_a$ rows of $\mathbf{H}'$ contain large values since the time delays of all sub-carriers fall into a certain period. We truncate the first $N_a$ rows of $\mathbf{H}'$ and denote the submatrix as $\mathbf{H}_a$. By feeding back $\mathbf{H}_a$ instead of $\mathbf{H}$, the overhead can be largely reduced.

However, $\mathbf{H}_a$ is still too heavy for uplink feedback since $N_t$ is a large number in massive MIMO system. Our purpose is to further compress matrix $\mathbf{H}_a$ to make the feedback as light as possible. The traditional CS based compressing algorithms rely on the sparsity of $\mathbf{H}_a$. But $\mathbf{H}_a$ is sparse enough only when $N_t \to\infty$, which is not possible in the real system \cite{wen2014channel}. By learning spatial patterns from the given CSI matrix, DL based encoder and decoder can compress and recover the matrix $\mathbf{H}_a$ with high precision even when it is dense.

The overview of the DL based feedback pipeline is demonstrated in Fig. \ref{System Model}. The frequency-delay domain CSI matrix $\mathbf{H}_a$ is first acquired with DFT and matrix truncation. The NN based ACRNet encoder compresses $\mathbf{H}_a$ into a $M$-dimensional feature vector $\mathbf{v}$. Therefore, the reciprocal compressive ratio $\eta$ can be derived as follows.
\begin{equation} \label{eq-eta}
  \eta = \frac{M}{2N_aN_t}
\end{equation}

Since the practical uplink feedback requires bitstream instead of float numbers, a quantizer is applied to generate binary feature $\mathbf{v}_b$ from original $\mathbf{v}$. The key feedback process at the UE can be described with (\ref{eq-encoder}).
\begin{equation} \label{eq-encoder}
  \mathbf{v}_b = \mathcal{Q}_\text{en}\left(\mathcal{E}_{\text{FC}}(\mathcal{E}_{\text{CNN}}(\mathbf{H}_a))\right),
\end{equation}
where $\mathcal{E}_{\text{CNN}}$ and $\mathcal{E}_{\text{FC}}$ are the CNN layers and the FC layer in ACRNet encoder, respectively. $\mathcal{Q}_\text{en}$ stands for the quantizer.

After the bitstream $\mathbf{v}_b$ is fed back to the BS, it is first dequantized into the float feature vector $\hat{\mathbf{v}}$ by a specialized dequantizer. Then $\hat{\mathbf{H}}_a$ is recovered from $\hat{\mathbf{v}}$ by ACRNet decoder and the $\hat{\mathbf{H}}$ is reconstructed via zero filling and inverse DFT. The core feedback process at the BS can be summarized as (\ref{eq-decoder}).
\begin{equation} \label{eq-decoder}
  \hat{\mathbf{H}}_a = \mathcal{D}_{\text{CNN}}(\mathcal{D}_{\text{FC}}(\mathcal{Q}_\text{de}(\mathbf{v}_b))),
\end{equation}
where $\mathcal{Q}_\text{de}$ is the dequantizer. $\mathcal{D}_{\text{FC}}$ and $\mathcal{D}_{\text{CNN}}$ represent the FC layer and CNN layers in ACRNet decoder, respectively.

% It is worth mentioning that the quantization and dequantization are actually optional modules when comparing the feedback network performance. Generally speaking, if a network achieves better feedback performance without quantization, its superiority is likely to remain after adding the quantization for practical deployment. The overall feedback process without quantization design can be described as follows.
% \begin{equation} \label{eq-overall}
%   \hat{\mathbf{H}}_a = \mathcal{D}_{\text{CNN}}(\mathcal{D}_{\text{FC}}(\mathcal{E}_{\text{FC}}(\mathcal{E}_{\text{CNN}}(\mathbf{H}_a)))),
% \end{equation}

Notably, uplink feedback is assumed to be perfect without noise. The CSI matrix is generated from COST2100 \cite{liu2012cost} channel model. Details of dataset generation settings are given in section \ref{Section5}.

\section{CSI Feedback with the proposed ACRNet} \label{Section3}

\subsection{The network aggregation aided ACRNet design} \label{Section3-1}

\begin{figure}[!b]
\centering
\includegraphics[width=0.55\textwidth]{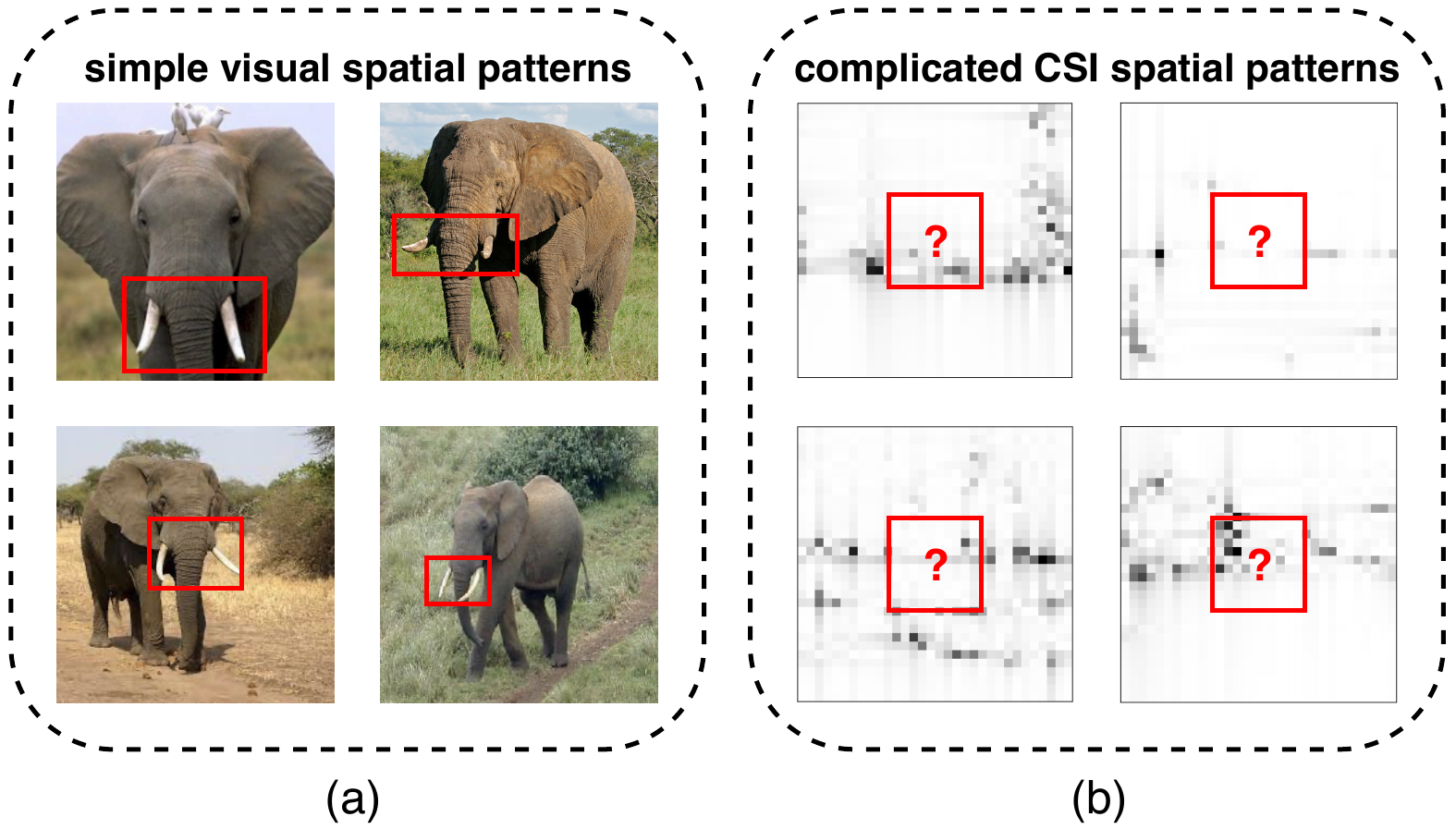}
\caption{Comparison of the possible spatial patterns between ordinary visual images and CSI matrices. As we can see, the ivory and the truck are common spatial patterns for different elephant images in (a). CNN is able to extract the specific feature by learning such patterns. However, the spatial patterns of CSI matrix in (b) is more complicated. A question mark is added since no explicit rules for CSI spatial patterns can be found.}
\label{CSI Diversity}
\end{figure}

The convolutional layer plays a significant role in modern neural network design. With the success of ResNet \cite{he2016deep}, the convolutional neural network (CNN) with the residual architecture has become the dominant design for many tasks including CSI feedback. One of the classic examples is the RefineNet proposed in CsiNet \cite{wen2018deep}, which is also used in later works like CsiNetPlus \cite{guo2020convolutional} and DS-NLCsiNet \cite{yu2020ds}. Therefore, the advanced design of convolutional layer is one of the key techniques for the feedback network enhancement. In fact, the vanilla convolutional layer is not the optimal choice since the diversity of the CSI spatial patterns is not considered well enough.

As it is shown in Fig. \ref{CSI Diversity}, the spatial pattern of CSI matrices is more complicated than the ordinary visual images. Take elephants in ImageNet \cite{deng2009imagenet} as an example, different samples of elephants share some common spatial patterns including the crooked ivory, the long truck and the scalloped ears. These common spatial patterns make it easier for the network to learn the useful feature from the image with vanilla convolution kernel.

At the mean while, three samples of the outdoor CSI matrices are shown in Fig. \ref{CSI Diversity}. Note that all the CSI matrices are visualized as $N_a\times N_t$ gray-scale images. The real and imaginary part of a CSI matrix is randomly chosen for visualization. As we can see, it is hard to find a common spatial pattern for the CSI matrices. In fact, the complexity of the CSI spatial patterns comes from the randomness of the multipath fading and the user locations.

However, the feedback network needs to find these complicated patterns in order to reduce the information loss of the CSI compression. The vanilla CNN is too weak to learn such diversified patterns. Therefore, more powerful CNN architecture design is needed for the CSI feedback task.

\begin{figure}[!t]
\centering
\includegraphics[width=0.55\textwidth]{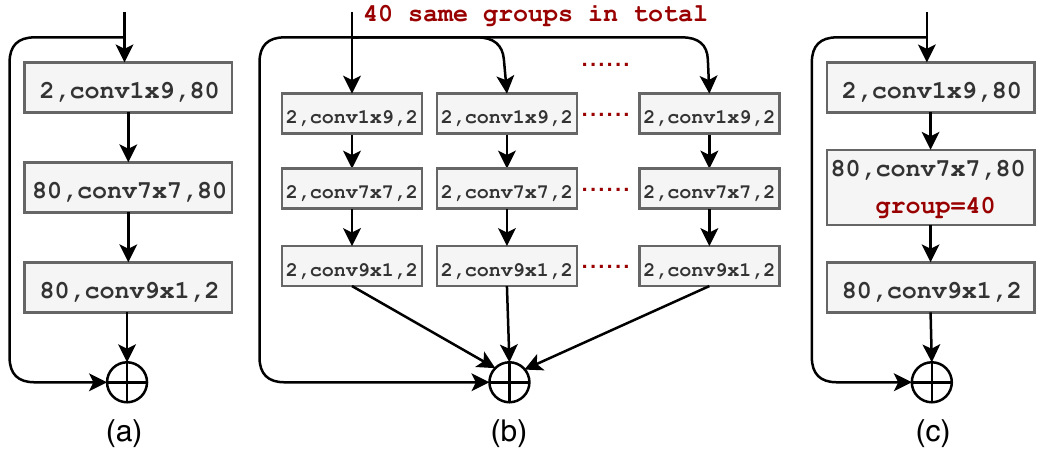}
\caption{Diagram of aggregated feedback block design for CSI feedback in ACRNet-10$\times$. The numbers at the beginning and the end of each convolutional layer stand for the dimension of input and output channels. (a) stands for the standard block with normal convolution, while (b) represents the corresponding aggregated feedback block in ACRNet-10$\times$ with 40 parallel groups. (c) is an equivalent realization of (b) with group convolution.}
\label{Group Convolution}
\end{figure}

In order to learn different CSI spatial patterns at the same time, we design a new feedback block based on the network aggregation technique \cite{xie2017aggregated}. As it is shown in Fig. \ref{Group Convolution}, the proposed aggregated feedback block contains dozens of parallel convolutional groups. Each group extracts relatively independent CSI feature so that the overall feature diversity is largely expanded.

However, the inference of the aggregated block in Fig. \ref{Group Convolution}-(b) is inefficient due to the massive parallel branches, which brings longer delay in the communication system. Luckily, the complicated aggregated block can be effectively realized with group convolution. Assume the dimensions of the input and output feature map at the $l^{th}$ layer are $D_{in}H_{in}W_{in}$ and $D_{out}H_{out}W_{out}$, respectively. The vanilla convolution at the $l^{th}$ layer can be expressed as follows:

\begin{equation} \label{eq-vanilla-conv}
  X_j^l = \sum_{i \in \mathbb{C}^l} X_i^{l-1} \circledast K_{ij}^l,
\end{equation}
where $X_i^{l-1}$ and $X_j^l$ are the $i^{th}$ and $j^{th}$ channel of the input and output feature map at the $l^{th}$ layer, respectively. $K_{ij}^l$ represents the $k\times k$ convolution kernel at the $l^{th}$ layer, which connects feature map $X_i^{l-1}$ and $X_j^l$. $\mathbb{C}^l$ is the set of input channel indices and $\circledast$ denotes the convolutional operation. Obviously, $|\mathbb{C}^l| = D_{in}$ for standard convolution and the corresponding floating point operations (FLOPs) is $D_{in}\times H_{out}W_{out}k^2 \times D_{out}$.

Compared with the vanilla convolution, the group convolution can be described as equation (\ref{eq-group-conv}) with similar notations.

\begin{equation} \label{eq-group-conv}
  X_j^l = \sum_{i \in \mathbb{G}^l_n} X_i^{l-1} \circledast K_{ij}^l,
\end{equation}
where the only difference compared with (\ref{eq-vanilla-conv}) is that the set of input channel indices $\mathbb{G}^l_n$ becomes a sub set of $\mathbb{C}^l$. The size of $\mathbb{G}^l_n$ is given in (\ref{eq-group-numbers}) when there is $g$ groups in total.

\begin{equation} \label{eq-group-numbers}
  \left|\mathbb{G}^l_n\right| = \frac{1}{g}D_{in} \;\;\;\;\text{for}\;\; \forall n \in \{1,\cdots,g\}
\end{equation}

It can be deduced from equation (\ref{eq-group-conv}) and (\ref{eq-group-numbers}) that each output channel in group convolution is only related with $D_{in}/g$ input channels. In other words, the convolutional operation is divided into $g$ disjoint parts, which is equivalent to the aggregation architecture with $g$ parallel groups in Fig. \ref{Group Convolution}-(b). Despite the equivalence, the group convolution scheme in Fig. \ref{Group Convolution}-(c) is more efficient on chip since it is optimized as an integrated operator in most of the deep learning frameworks like PyTorch and TensorFlow.

\begin{figure}[!t]
\centering
\includegraphics[width=0.45\textwidth]{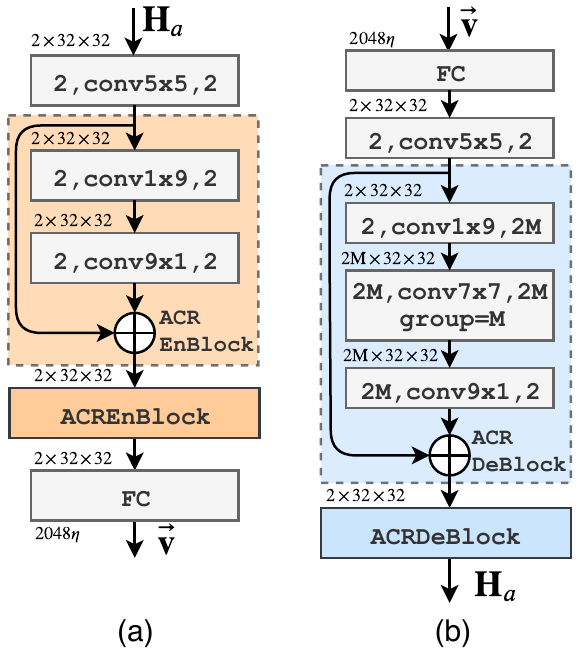}
\caption{The proposed ACRNet architecture. (a) and (b) present the encoder and decoder of ACRNet, respectively. The shape of input feature map $c\times h\times w$ is given on the top of each block. Note that each convolutional layer is followed by a batch normalization (BN) layer, and all activation functions and reshape layers are omitted for simplicity.}
\label{ACRNet Architecture}
\end{figure}

Therefore, we choose to implement the aggregated feedback block based on the group convolution as Fig. \ref{Group Convolution}-(c) shows. The kernel size is set to $7\times 7$ and the number of channels in each group is always two, aligning with the original CSI matrix. Convolution factorization is adopted as \cite{lu2020multi}, splitting a standard $9 \times 9$ convolution kernel into a $1 \times 9$ kernel and a $9 \times 1$ kernel. The factorization largely reduces the computational complexity while keeping the receptive field. In addition, residual architecture is adopted to preserve the spatial information as RefineNet \cite{wen2018deep} and CRBlock \cite{lu2020multi}.

With the aforementioned aggregated feedback block, we design a novel network named ACRNet as Fig. \ref{ACRNet Architecture} shows. For ACRNet encoder, the CNN layers consist of a $5\times 5$ convolution head and a pair of ACREnBlocks. The FC layer compresses the extracted feature to the target dimension $2048\eta$, where $\eta$ is the reciprocal compression ratio in (\ref{eq-eta}). For ACRNet decoder, the FC layer restores the reduced dimension and the following CNN layers reconstruct the original CSI spatial pattern.

The key design to the decoder CNN layers is the ACRDeBlock, which enriches the diversity of the learned feature efficiently. As we can see, the ACRDeBlock is an elastic architecture and the aggregated block in Fig. \ref{Group Convolution}-(c) is an instance when $M=40$. The reasonability of the elastic network design will be explained in section \ref{Section4-1}. Notably, the details of the activation function design are given in the next section, therefore it is omitted in Fig. \ref{ACRNet Architecture} for a clearer view.

\subsection{The learnable parametric ReLU activation} \label{Section3-2}

In this section, we will explain the activation design of the proposed ACRNet. Note that the activation layers discussed below do not include the last one, which is always sigmoid function for output range normalization.

Traditional sigmoid and tanh activation tend to squash the gradient since their output is limited in a small range. As a result, the gradient becomes unacceptably small as the network goes deeper. The rectified linear unit (ReLU) activation in (\ref{eq-relu}) is adopted by most of the mainstream networks like ResNet. It avoids the gradient vanishing while keeping the nonlinearity.

\begin{equation} \label{eq-relu}
	\ReLU(x) = \left\{
		\begin{array}{ll}
			x, & x \ge 0 \\
			0, & x < 0
		\end{array}
	\right.
\end{equation}

However, the zero forcing of the negative gradient in ReLU might cause some information loss. The loss is negligible for huge networks like ResNet50, but the feedback networks are usually light for practical deployment and real time inference. Therefore, leaky ReLU (LReLU) in (\ref{eq-lrelu}) is introduced to improve the feedback network ability in CsiNet \cite{wen2018deep}.

\begin{equation} \label{eq-lrelu}
	\LReLU(x) = \left\{
		\begin{array}{ll}
			x, & x \ge 0 \\
			Cx, & x < 0,
		\end{array}
	\right.
\end{equation}
where $C \in \mathbb{R}^+$ is a given const, representing the negative slop. Experiments in CRNet \cite{lu2020multi} shows that better feedback performance can be achieved when $C=0.3$.

For our proposed ACRNet, even the LReLU becomes inadequate. As it is mentioned in section \ref{Section3-1}, the key advantage of the aggregated block is that the feature extracted by each convolutional group is relatively independent. Large number of aggregated groups provide sufficient degree of freedom for the network so that the diversified CSI feature can be better learned. However, the negative slopes for all the groups are the same under LReLU activation, which limits such degree of freedom and harms the feedback performance.

In order to solve the challenge above, we introduce parametric ReLU (PReLU) to the CSI feedback task. The negative slope on each convolutional channel is a learnable variable instead of a fixed const. This offers higher degree of freedom, which matches well with the aggregated architecture of ACRNet. The PReLU activation can be expressed as follows.

\begin{equation} \label{eq6}
	\PReLU(x) = \left\{
		\begin{array}{ll}
			x, & x \ge 0 \\
			\alpha x, & x < 0,
		\end{array}
	\right.
\end{equation}
where $\alpha \in \mathbb{R}$ is the learnable negative slope on the corresponding channel. Note that the number of different $\alpha$ is determined by the number of input channels.

\begin{figure}[!t]
\centering
\includegraphics[width=0.55\textwidth]{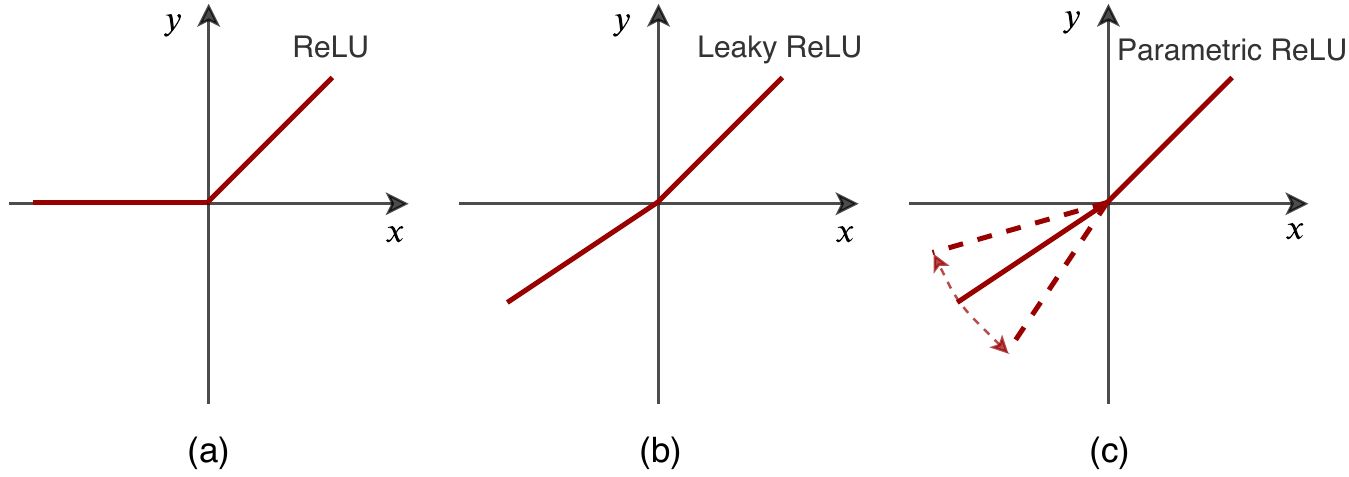}
\caption{Comparison of ReLU, leaky ReLU and parametric ReLU activation.}
\label{Activation}
\end{figure}

The comparison among vanilla ReLU, LReLU and PReLU is shown in Fig. \ref{Activation}. Obviously, the PReLU activation is a generalization of the original LReLU. If we fix all the $\alpha$ to $0.3$, the PReLU will degenerate to the LReLU in previous works. Therefore, the PReLU should be more powerful compared with LReLU if all the $\alpha$ are properly trained.

\begin{figure}[!b]
\centering
\includegraphics[width=\textwidth]{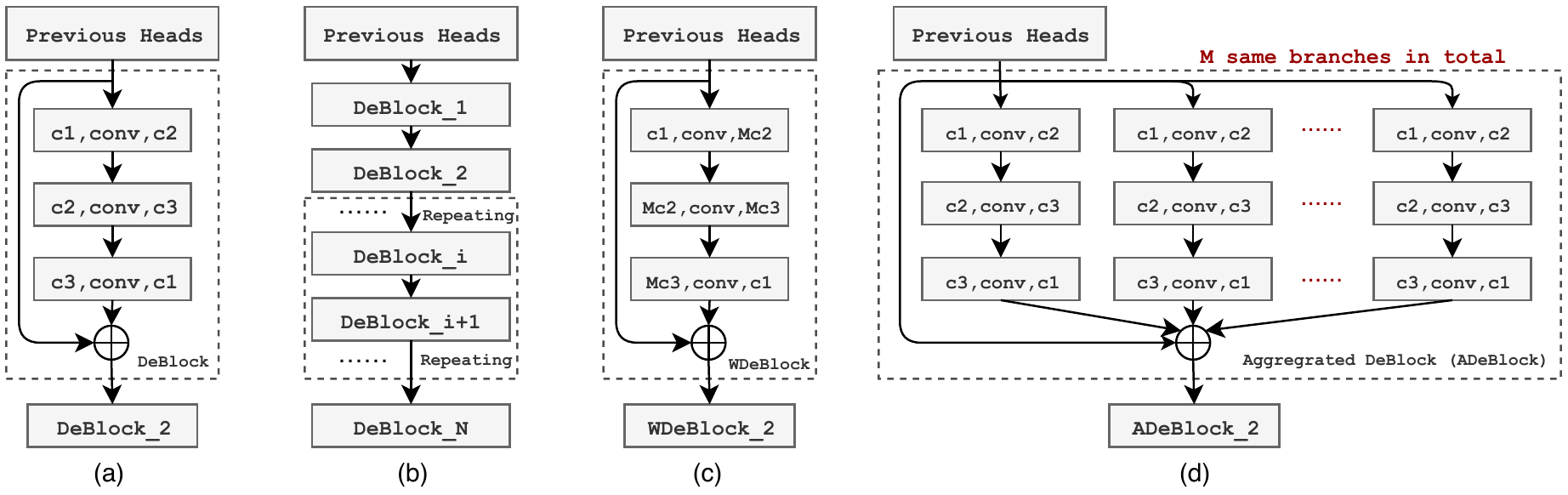}
\caption{General methods of network expansion. The digits at the beginning and the end of each convolutional layer represent the number of input and output channels. The expansion focus on the decoder at the BS since the encoder is resource sensitive. (a) is the original decoder to be expanded. (b) illustrates the depth expansion while (c) explains the width expansion. Finally (d) shows the width expansion on aggregated block, which is used in the proposed ACRNet.}
\label{Network Expansion}
\end{figure}

\section{Flexible DL Deployment for CSI Feedback} \label{Section4}

\subsection{The elastic architecture design in ACRNet} \label{Section4-1}

In this section, we design a practical network expansion scheme based on ACRNet to solve the elastic deployment challenge for CSI feedback. In the real communication system, the hardware resources allocated to the feedback module are not fixed. Therefore, elastic architecture design that can adapt to different resource limitations is necessary for the practical feedback network deployment.

For example, the network should be effectively expanded in exchange for better feedback precision when there are extra computational resources. However, the complexities of the existing feedback networks are usually fixed, which can not meet such demand.

In order to design the practical elastic network, we need to look into the common ways of network expansion. As it is shown in Fig. \ref{Network Expansion}, there are two basic dimensions that can expand a given network, the depth and the width. The depth expansion adds extra blocks in a serial way as it is presented in Fig. \ref{Network Expansion}-(b). With the help of residual architecture, extremely deep network becomes very powerful for task like classification and object detection \cite{he2016deep}. On the other hand, the width expansion enlarges the number of channels in the existing blocks as Fig. \ref{Network Expansion}-(c) shows. Larger $M$ gives wider convolutional channels and the network ability is enhanced with the quadratic growth of complexity. One of the typical examples of width expansion is the Wide ResNet (WRN) proposed in \cite{sergey2016wide}, which improves the network performance with a wide and shallow structure.

\begin{figure}[!b]
\centering
\includegraphics[width=0.5\textwidth]{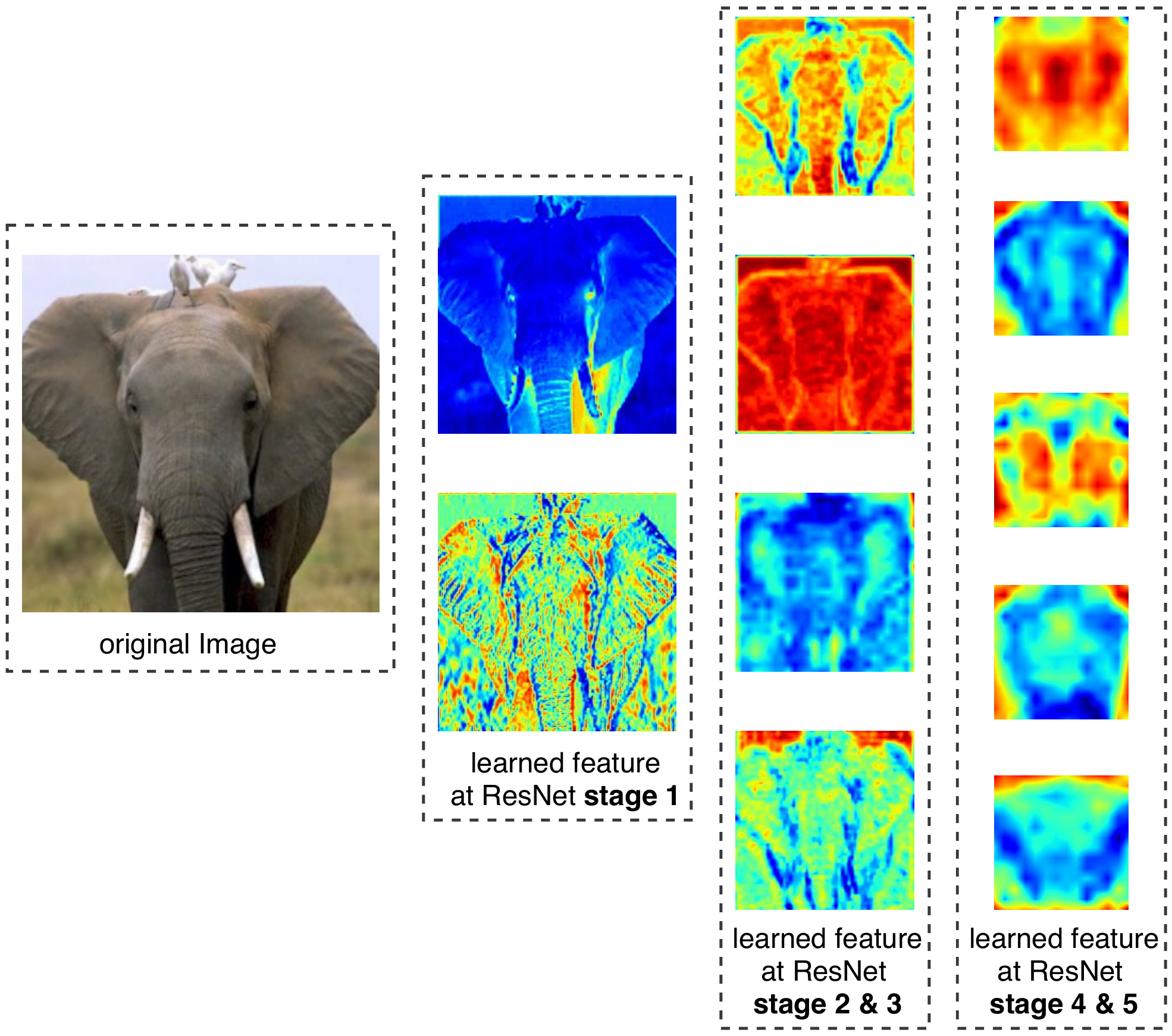}
\caption{Comparison of the learned spatial patterns from a ResNet style network. As the network goes deeper, the spatial information becomes vaguer and the specific spatial patterns are lost. An elephant image from ImageNet \cite{deng2009imagenet} is chosen as an example since the spatial patterns of the CSI matrix are not clear enough to view.}
\label{Spatial Information Loss}
\end{figure}

When it comes to the CSI feedback task, we argue that width expansion is more suitable than depth expansion. The feedback network is actually an auto encoder for unsupervised learning. The purpose of the network is to compress and recover the spatial information of the given CSI matrix. Therefore, richer spatial information is essential to better CSI feedback.

At the mean while, it is widely known that the semantic information becomes richer while the spatial information becomes vaguer as the network goes deeper. A perceptual intuition is given in Fig. \ref{Spatial Information Loss} for a clearer view. As we can see, the spatial patterns like the shape of the trunk are highly preserved at the earlier stage of the network. However, these patterns are totally lost when the network becomes deeper at stage 4 and stage 5 of the ResNet style network.

In order to preserve the crucial spatial information, the depth of the feedback network should be limited. Therefore, the width expansion is more appropriate for the elastic feedback network design. Specifically, width of the aggregated block can be conveniently expanded by increasing the number of groups as Fig. \ref{Network Expansion}-(d) shows.

For the proposed elastic ACRNet in Fig. \ref{ACRNet Architecture}, users can simply expand the network by increasing $M$ in the ACRDeBlock. For instance, The lightest ACRNet-1$\times$ sets $M$ to $4$, while the polyploid ACRNet-$k\times$ amplifies the factor $M$ to $4k$. In this way, the proper width expansion brings better feedback performance at the cost of heavier decoder. The resource sensitive encoder at the UE remains unchanged, which is more practical for the systematic deployment.

It is worth mentioning that joint optimization of width, depth, kernel size and multi-resolution branches can make the feedback network even more powerful. However, such optimization is so complicated that it needs the help of neural architecture search (NAS) \cite{barret2017neural} technique. The combination of NAS and feedback network design might be an interesting research tendency in the future.

\subsection{The FC binarization and feature quantization} \label{Section4-2}

In this section, the network binarization is combined with vector quantization in ACRNet to reduce the deployment cost and the feedback overhead. With the help of the elastic architecture design and the binarization with quantization technique, a comprehensive strategy for flexible deployment is proposed for ACRNet based CSI feedback.

\begin{figure}[!t]
\centering
\includegraphics[width=0.55\textwidth]{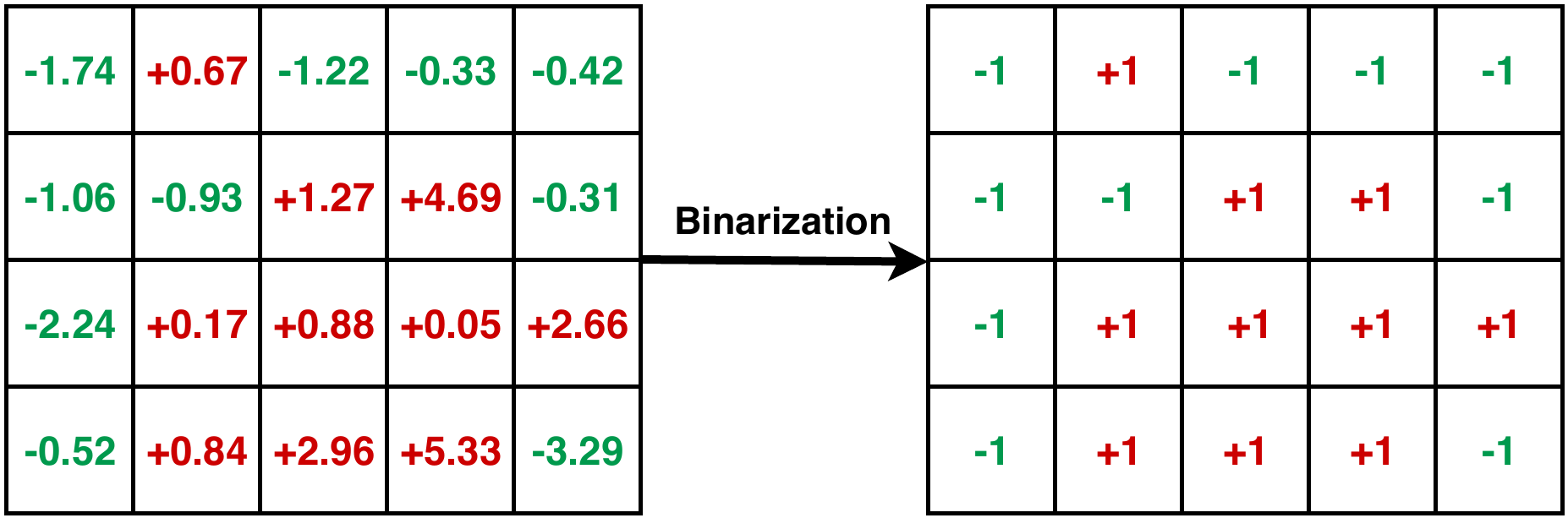}
\caption{A demonstration of FC weight binarization. The full precision float numbers are binarized based on their sign.}
\label{FC Binarization}
\end{figure}

Network binarization is a technique to design extremely lightweight network \cite{rastegari2016xnor}. As demonstrated in Fig. \ref{FC Binarization}, the network weights change from the 32-bit float number to the 1-bit binary number. As it is analyzed in \cite{lu2021binary}, the FC layer is the main contributor to the computation and memory cost of the mainstream feedback encoder. Specifically, the FC layer occupies 80.76\% of the computational complexity and 99.98\% of the memory cost for ACRNet encoder. By following the algorithm proposed in \cite{lu2021binary}, over 30$\times$ memory saving and around 2$\times$ inference acceleration can be obtained at the resource sensitive UE.

Notably, the feedback decoder still adopts a full-precision FC layer in \cite{lu2021binary}. However, the FC layer takes up over 96\% of the decoder memory cost for even the heaviest ACRNet-20$\times$. Therefore, it would be better if the decoder FC layer is also binarized, which is proved to be feasible based on the ACRNet design. In this way, both the encoder and the decoder can enjoy a huge memory saving.

Furthermore, the aforementioned binarization scheme is combined with vector quantization scheme for the first time. The quantization is necessary for the practical deployment since the uplink digital feedback can only send the bitstream. And the feedback overhead is unacceptable if the feature vector $\mathbf{v}$ is directly fed back as the 32-bit float number. We adopt a $B$-bit uniform quantization as follows.
\begin{equation} \label{eq-quant}
    x_b = \left\lfloor x\cdot2^B \right\rfloor,\;\;x\in(0,1),
\end{equation}
where $x_b$ and $x$ are the corresponding elements of $\mathbf{v}_b$ and $\mathbf{v}$. All the input elements $x$ are normalized to $(0,1)$ by a sigmoid function. With the feature vector $\mathbf{v}$ quantized, the feedback overhead is reduced for $32/B$ times. Note that the quantization and dequantization are not derivable. Their gradients are fixed to 1 as \cite{lu2019bit} and \cite{guo2020convolutional} so that the the quantized network can be normally trained in an end-to-end way.

\begin{algorithm}[!t]
\renewcommand{\algorithmicrequire}{\textbf{Input:}}
\renewcommand\algorithmicensure {\textbf{Output:}}
\caption{The flexible deployment strategy of ACRNet}
\label{algorithm-deployment}
\begin{algorithmic}[1]

\REQUIRE The resource limitation of the UE and BS. The upper limit of the feedback bits $\max(N_{fb})$.
\ENSURE The deployment scheme of ACRNet
\STATE Choose the largest $M$ for the aggregated block within the resource limitation at the BS.
\STATE Binarize the FC layer if the memory is insufficient at the UE or BS.
\STATE Choose a proper combination of $\eta$ and $B$ based equation (\ref{eq-quant-trade-off}) and the feedback overhead limitation $\max(N_{fb})$.
\STATE Give the comprehensive deployment strategy based on the previous choices. Train and deploy the network as usual.

\end{algorithmic}
\end{algorithm}

It is obvious that the bit-wise feedback overhead is determined by both the compression ratio and the quantization precision for a certain feedback system. The number of the feedback bits $N_{fb}$ is given as follows.

\begin{equation} \label{eq-quant-trade-off}
  N_{fb} = 2N_aN_t\times\eta\times B
\end{equation}

In conclusion, three techniques for the practical deployment of ACRNet is proposed in section \ref{Section4} including the elastic architecture design, the FC binarization and the feature vector quantization. These techniques can be flexibly combined to provide a best practice based on the condition of the real communication system. The flexible deployment strategy of the proposed ACRNet is summarized in Algorithm \ref{algorithm-deployment}.

\section{Simulation Results and Analysis} \label{Section5}

\subsection{Experimental settings} \label{Section5-1}

As many previous works, we follow the experimental settings used in CsiNet \cite{wen2018deep}. Based on COST2100 channel model \cite{liu2012cost}, the indoor scenario at 5.3GHz and the outdoor scenario at 300MHz are considered. FDD system with 1024 sub-carriers is adopted and $N_a$ is set to 32. For massive MIMO system, uniform linear array (ULA) model is used with $N_t=$ 32. The training and test dataset are independently generated with 100,000 and 20,000 CSI matrices, respectively.

As for the deep learning hyper-parameters, the batch size is 200 and the loss function is the mean square error (MSE) loss. Adam optimizer is used with $\beta_1=$ 0.9 and $\beta_2=$ 0.999. We adopt the warm up aided cosine annealing scheduler introduced in \cite{lu2020multi}, which can be derived as follows.

\begin{equation}
  \gamma = \gamma_{min} + \frac{1}{2}\left(\gamma_{max} - \gamma_{min}\right)\left(1 + \cos \left( \frac{t-T_w}{T-T_w} \pi \right)\right)_,
\end{equation}
where $t$ stands for the index of current epoch and $\gamma$ is the corresponding learning rate. The number of training epochs $T=$ 2500 and the number of warm up epochs $T_w=$ 30. The initial learning rate $\gamma_{max}=4\times 10^{-3}$ while the minimal learning rate $\gamma_{min}=5\times 10^{-5}$.

\begin{table}[H]
\caption{NMSE (dB) and complexity comparison between series of feedback networks and proposed elastic ACRNet}
\begin{center}
\makegapedcells \renewcommand\tabcolsep{10pt}
\begin{tabular}{c l | r r | r r}
\Xhline{0.8pt}
\multirow{2}{*}{$\mathbf{\eta}$} & \multirow{2}{*}{\textbf{Methods}} & \multicolumn{2}{c|}{\textbf{Complexity}} & \multicolumn{2}{c}{\textbf{NMSE}$^{\mathrm{a}}$} \\
& & \multicolumn{1}{c}{FLOPs} & \multicolumn{1}{c|}{params} & \multicolumn{1}{c}{indoor} & \multicolumn{1}{c}{outdoor} \\
\Xhline{0.8pt}
\multirow{7}{*}{1/4}
  & CsiNet\cite{wen2018deep} & 5.41M & 2103K & -17.36 & -8.75 \\
  & CRNet\cite{lu2020multi} & 5.12M & 2103K & -26.99 & -12.70 \\
  & DS-NLCsiNet\cite{yu2020ds} & 11.30M & 2108K & -24.99 & -12.09 \\
  & CsiNetPlus\cite{guo2020convolutional} & 24.57M & 2122K & -27.37 & -12.40 \\
  & ACRNet-1$\times$ & \textbf{4.64M} & \textbf{2102K} & -27.16 & -10.71 \\
  & ACRNet-10$\times$ & 24.40M & 2123K & -29.83 & -13.61 \\
  & ACRNet-20$\times$ & 46.36M & 2145K & \textbf{-32.02} & \textbf{-14.25} \\
\hline
\multirow{6}{*}{1/8}
  & CsiNet\cite{wen2018deep} & 4.37M & 1054K & -12.70 & -7.61 \\
  & CRNet\cite{lu2020multi} & 4.07M & 1054K & -16.01 & -8.04 \\
  & DS-NLCsiNet\cite{yu2020ds} & 10.25M & 1059K & -17.00 & -7.96 \\
  & CsiNetPlus\cite{guo2020convolutional} & 23.52M & 1073K & -18.29 & -8.72 \\
  & ACRNet-1$\times$ & \textbf{3.60M} & \textbf{1054K} & -15.34 & -7.85 \\
  & ACRNet-10$\times$ & 23.36M & 1074K & -19.75 & -9.22 \\
  & ACRNet-20$\times$ & 45.31M & 1096K & \textbf{-20.78} & \textbf{-9.68} \\
\hline
\multirow{6}{*}{1/16}
  & CsiNet\cite{wen2018deep} & 3.84M & 530K & -8.65 & -4.51 \\
  & CRNet\cite{lu2020multi} & 3.55M & 530K & -11.35 & -5.44 \\
  & DS-NLCsiNet\cite{yu2020ds} & 9.72M & 534K & -12.93 & -4.98 \\
  & CsiNetPlus\cite{guo2020convolutional} & 23.00M & 549K & -14.14 & -5.73 \\
  & ACRNet-1$\times$ & \textbf{3.07M} & \textbf{529K} & -10.36 & -5.19 \\
  & ACRNet-10$\times$ & 22.83M & 549K & -14.32 & -6.30 \\
  & ACRNet-20$\times$ & 44.79M & 572K & \textbf{-15.05} & \textbf{-6.47} \\
\hline
\multirow{6}{*}{1/32}
  & CsiNet\cite{wen2018deep} & 3.58M & 268K & -6.24 & -2.81 \\
  & CRNet\cite{lu2020multi} & 3.28M & 267K & -8.93 & -3.51 \\
  & DS-NLCsiNet\cite{yu2020ds} & 9.46M & 272K & -8.64 & -3.35 \\
  & CsiNetPlus\cite{guo2020convolutional} & 22.74M & 286K & -10.43 & -3.4 \\
  & ACRNet-1$\times$ & \textbf{2.81M} & \textbf{267K} & -8.60 & -3.31 \\
  & ACRNet-10$\times$ & 22.57M & 287K & -10.52 & -3.83 \\
  & ACRNet-20$\times$ & 44.52M & 309K & \textbf{-10.77} & \textbf{-4.05} \\
\hline
\multirow{6}{*}{1/64}
  & CsiNet\cite{wen2018deep} & 3.45M & 137K & -5.84 & -1.93 \\
  & CRNet\cite{lu2020multi} & 3.16M & 136K & -6.49 & -2.22 \\
  & DS-NLCsiNet\cite{yu2020ds} & 9.33M & 141K & / & / \\
  & CsiNetPlus\cite{guo2020convolutional} & 22.61M & 155K & / & / \\
  & ACRNet-1$\times$ & \textbf{2.68M} & \textbf{136K} & -6.51 & -2.29 \\
  & ACRNet-10$\times$ & 22.44M & 156K & -7.44 & -2.61 \\
  & ACRNet-20$\times$ & 44.39M & 178K & \textbf{-7.78} & \textbf{-2.69} \\
\Xhline{0.8pt}
\multicolumn{5}{l}{$^{\mathrm{a}}$ / means the performance is not reported.} \\
\end{tabular}
\label{tab1}
\end{center}
\end{table}

\subsection{Performance of the proposed ACRNet} \label{Section5-2}

Generally, the performance of CSI feedback is measured by the normalized mean square error (NMSE) between the original angular-delay domain CSI matrix $\mathbf{H}_a$ and the recovered $\hat{\mathbf{H}}_a$.

\begin{equation} \label{eq8}
	\text{NMSE} = \mathbb{E}\left\{\frac{\Vert \mathbf{H}_a - \hat{\mathbf{H}}_a \Vert_2^2} {\left\Vert \mathbf{H}_a\right\Vert_2^2} \right\}
\end{equation}

Table \ref{tab1} shows the NMSE performance comparison among the proposed ACRNet and series of influential previous feedback networks. According to the results reported in \cite{wen2018deep}, the feedback NMSE of CsiNet is much better than traditional compressed sensing algorithms. Therefore it is unnecessary to list those compressed sensing based schemes here.

As we can see, the basic ACRNet-1$\times$ is the lightest among all these networks, yet it completely outperforms the heavier CsiNet with decent CSI reconstruction results. In fact, the lightweight ACRNet-1$\times$ even achieves better performance under some specific situation compared with the advanced DS-NLCsiNet. For instance, the NMSE of ACRNet-1$\times$ is 2.17 dB lower than DS-NLCsiNet for indoor scenario with reciprocal compression ratio $\eta=1/4$. Note that DS-NLCsiNet is an advanced feedback network based on the powerful non-local architecture. This proves the effectiveness of the prototypical design of ACRNet based on the network aggregation.

\begin{figure}[!b]
\centering
\includegraphics[width=0.55\textwidth]{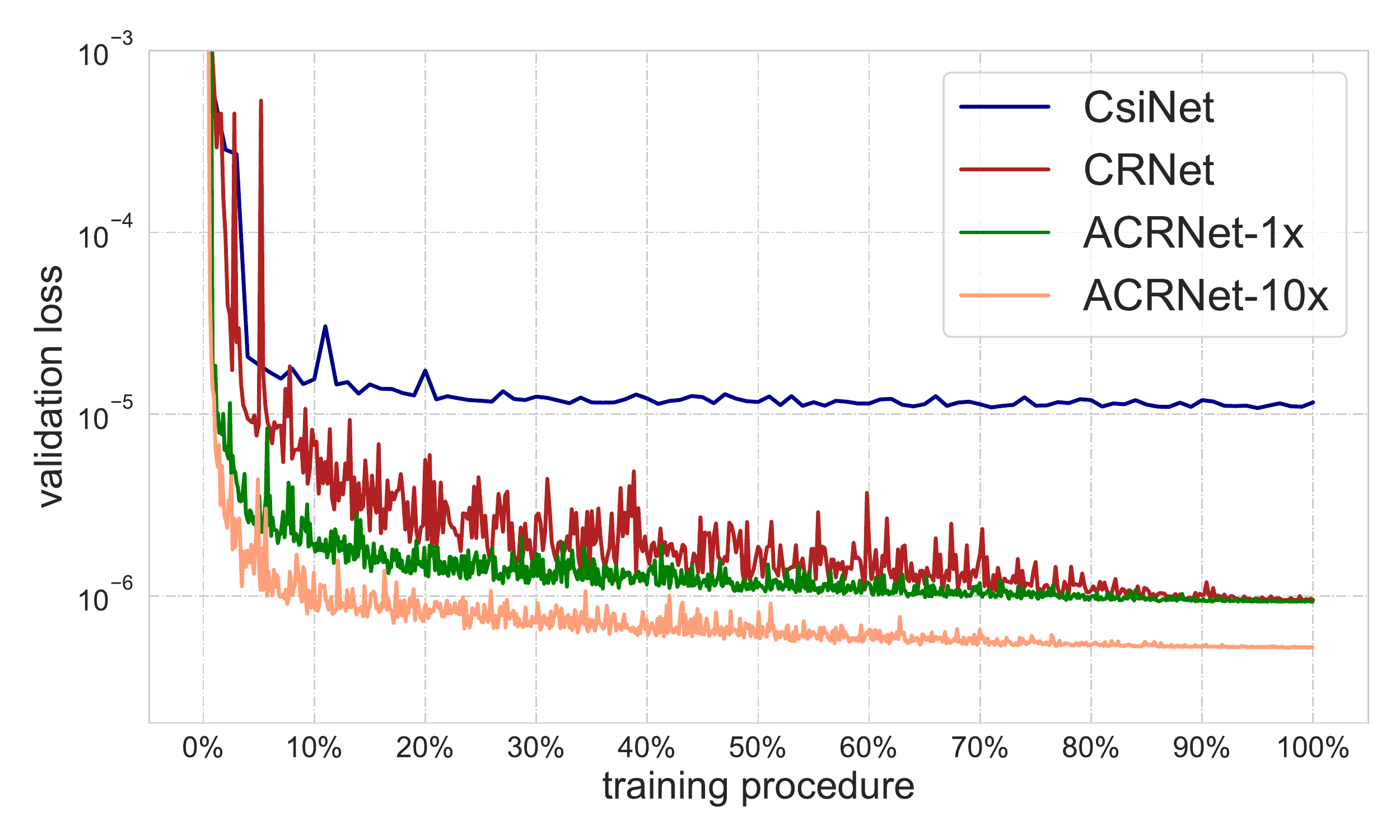}
\caption{The comparison of validation loss among two representative benchmark networks (CsiNet and CRNet) and the proposed ACRNet. The convergence of the proposed ACRNet-1$\times$ is more stable than CRNet with similar performance.}
\label{Table1 Loss}
\end{figure}

The performance of heavier ACRNet is also impressive. ACRNet-10$\times$ outperforms the CsiNetPlus with less computational complexity under all the compression ratios for both indoor and outdoor scenario. In addition, ACRNet-20$\times$ achieves state-of-the-art feedback performance when the FLOPs complexity is limited within 50M. For example, the feedback NMSE of indoor scenario is reduced to -32 dB for the first time when $\eta=1/4$.

It can be observed that the performance improvement is more statistically apparent for lower compression ratio. Based on the NMSE performance of CsiNetPlus, the proposed ACRNet-10$\times$ achieves 2.46 dB and 1.21 dB improvement for $\eta=1/4$ under indoor and outdoor scenario, respectively. The gaps of NMSE drop to 0.12 dB and 0.43 dB for $\eta=1/32$, which indicates that CSI reconstruction under high compression ratio is much harder.

Another crucial aspect of network design is the convergence. As it is demonstrated in Fig. \ref{Table1 Loss}, the validation MSE loss of the proposed ACRNet decays rapidly at the beginning of the training and remains stable later on, which shows satisfactory convergence capacity.

\begin{figure}[!t]
\centering
\includegraphics[width=0.55\textwidth]{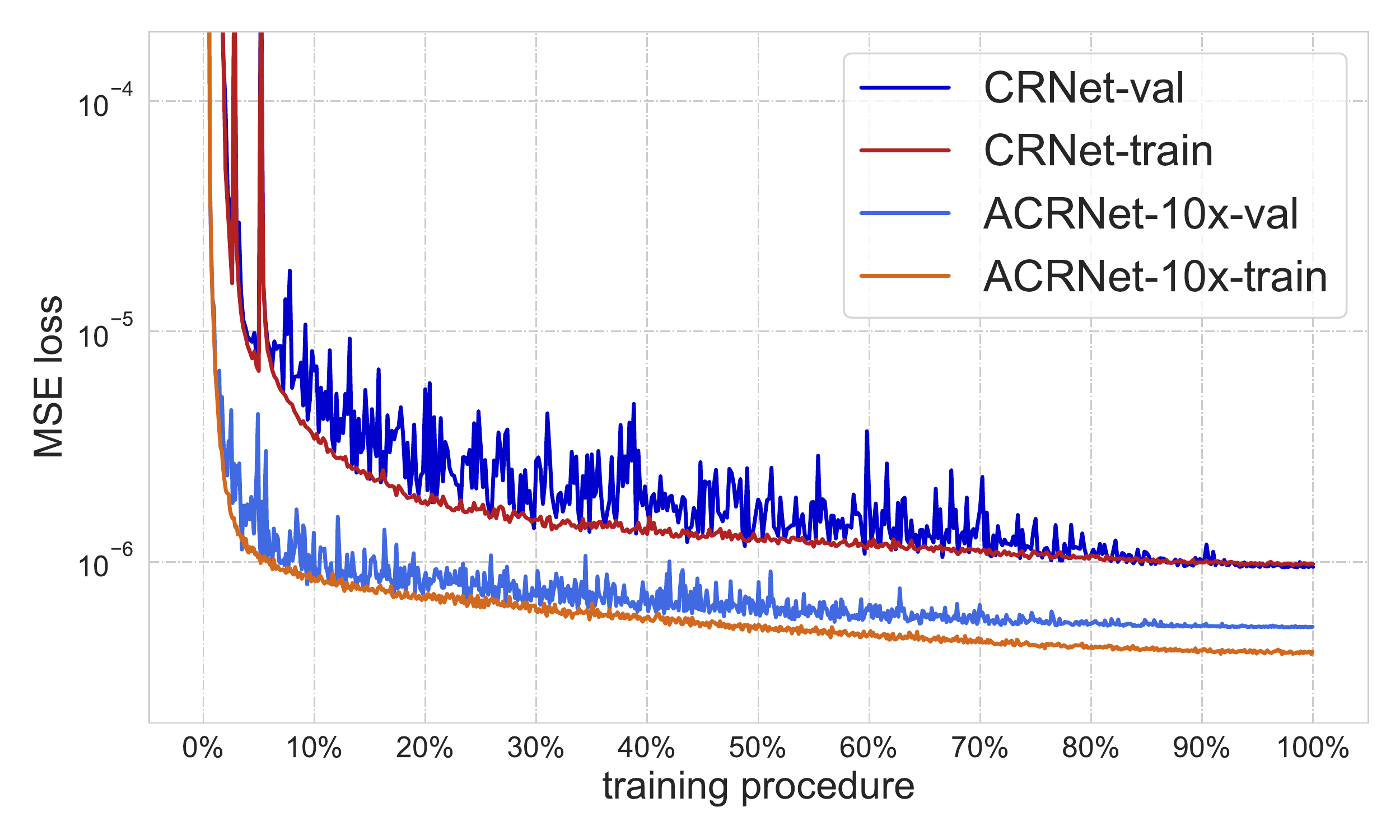}
\caption{The descending tendency comparison of validation and training loss of CRNet and the proposed ACRNet-10$\times$. The overfitting of ACRNet-10$\times$ is larger but acceptable.}
\label{Table1 Overfitting}
\end{figure}

In the end, the validation and training loss of CRNet and ACRNet-10$\times$ are compared in Fig. \ref{Table1 Overfitting} to measure the overfitting. It is obvious that the overfitting of ACRNet-10$\times$ is more severe than CRNet since the gap between training and validation loss is larger. However, such overfitting is acceptable since the validation loss is still decreasing instead of increasing as the training loss declines.

\subsection{Effectiveness of PReLU activation} \label{Section5-3}

In this subsection, some ablation studies are presented to prove the effectiveness of PReLU activation in CSI feedback task. In addition, the distributions of learned negative slope $\alpha$ in (\ref{eq6}) are analyzed based on different ACRNet variants.

\begin{table}[!t]
\caption{NMSE (dB) comparison on different activation functions}
\begin{center}
\makegapedcells \renewcommand\tabcolsep{10pt}
\begin{tabular}{c c | r r r}
\Xhline{0.8pt}
\multirow{2}{*}{$\eta$} & \multirow{2}{*}{\textbf{Scenario}}
  & \multicolumn{3}{c}{\textbf{Activation Scheme}} \\
& & LReLU & SPReLU & PReLU \\
\Xhline{0.8pt}
\multirow{2}{*}{1/4}
  & indoor & -24.25 & -25.88 & \textbf{-27.16} \\
  & outdoor & -10.13 & -10.25 & \textbf{-10.71} \\
\hline
\multirow{2}{*}{1/8}
  & indoor & -14.24 & -14.17 & \textbf{-15.34} \\
  & outdoor & -7.17 & -7.21 & \textbf{-7.85} \\
\hline
\multirow{2}{*}{1/16}
  & indoor & -9.83 & -10.02 & \textbf{-10.36} \\
  & outdoor & -4.98 & -5.12 & \textbf{-5.19} \\
\hline
\multirow{2}{*}{1/32}
  & indoor & -8.37 & -8.49 & \textbf{-8.60} \\
  & outdoor & -3.03 & -3.22 & \textbf{-3.31} \\
\hline
\multirow{2}{*}{1/64}
  & indoor & -5.67 & -6.29 & \textbf{-6.51} \\
  & outdoor & -2.13 & -2.11 & \textbf{-2.29} \\
\hline
\Xhline{0.8pt}
\end{tabular}
\label{tab3}
\end{center}
\end{table}

Two different types of parametric ReLU are considered in the ablation study. The solely parametric ReLU (SPReLU) learns only one $\alpha$ for the whole activation layer. At the mean while, the parametric ReLU (PReLU) learns different $\alpha$ for each output channel of the preceding convolutional layer. The vanilla ReLU is too weak for CSI feedback, so we only test the performance of LReLU in the ablation study.

We train ACRNet-1$\times$ based on the LReLU, SPReLU and PReLU with different compression ratio at both indoor and outdoor scenario. The $\alpha$ is initialized to 0.3 and updated during training. The NMSE performances of different activation functions are given in Table \ref{tab3}. As we can see, SPReLU tends to work better than the original LReLU while the PReLU outperforms both of them under all circumstances. Therefore the extra degree of freedom provided by the learnable $\alpha$ does help the CSI feature extraction of ACRNet. Experiments in Table \ref{tab3} show that SPReLU with layer-wise learnable $\alpha$ is not enough. The parallel architecture in ACRNet requires higher degree of freedom on each convolutional group. That is why PReLU with channel-wise learnable $\alpha$ is adopted.

\begin{figure}[!t]
\centering
\includegraphics[width=0.55\textwidth]{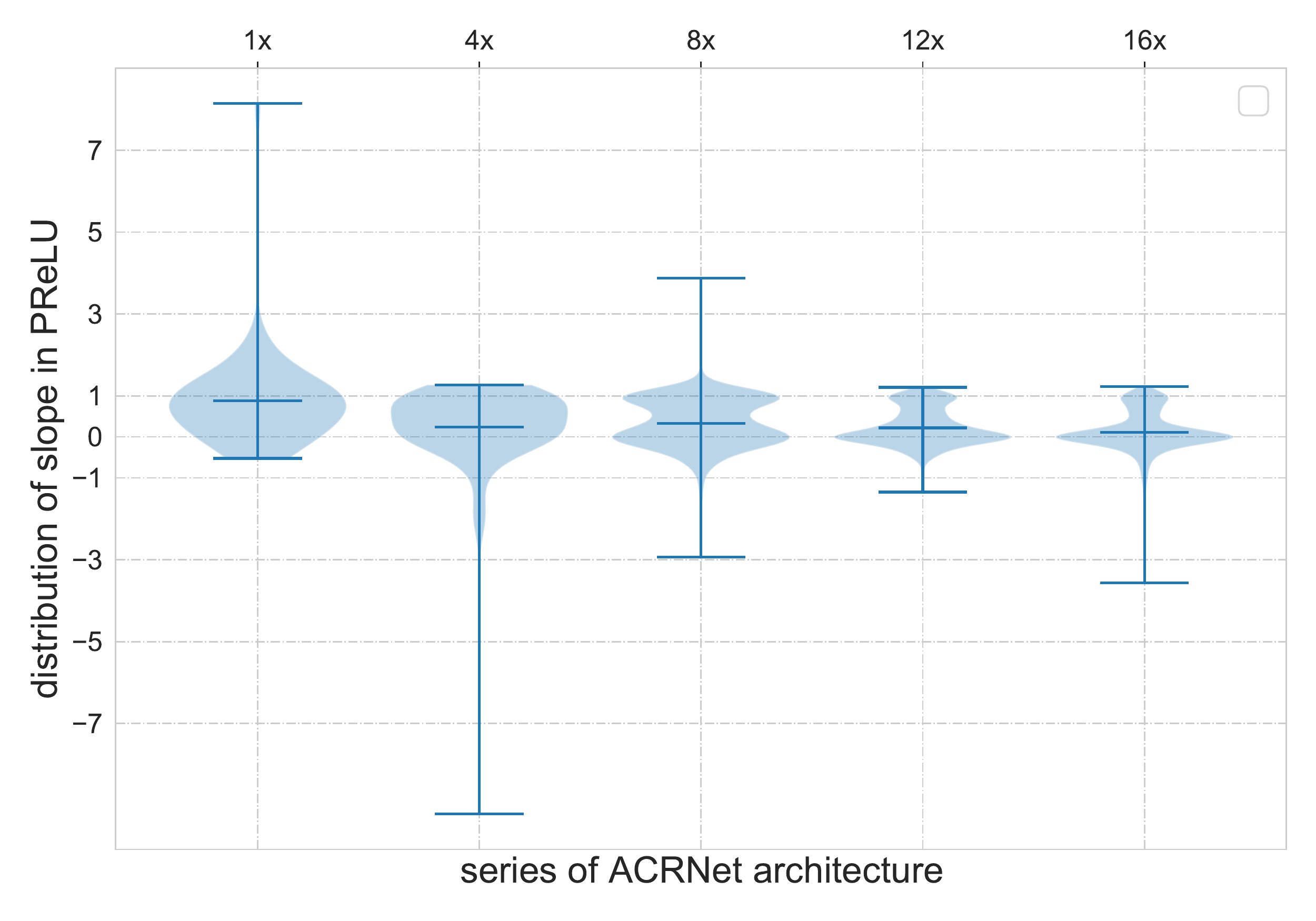}
\caption{Distribution graph of the learned slope of PReLU for ACRNet variants. The three parallel lines represent the maximum, minimum and mean of the learned $\alpha$. The shaded part shows the shape of probability distribution function (PDF) of $\alpha$ for each variant.}
\label{Slope Distribution}
\end{figure}

The distributions of the best learned $\alpha$ in several ACRNet variants are further studied in Fig. \ref{Slope Distribution}. The mean of all the learned $\alpha$ in ACRNet tends to be smaller as the expansion multiple goes higher. In addition, local maxima of the distribution appears at $\alpha=1$ and $\alpha=0$ for heavier ACRNet, which stand for absolute activation and vanilla ReLU activation, respectively. The combination of these two activation functions might be a better choice without extra cost from the learnable parameters.

\subsection{Effectiveness of the elastic ACRNet design} \label{Section5-4}

\begin{figure}[!b]
\centering
\includegraphics[width=0.55\textwidth]{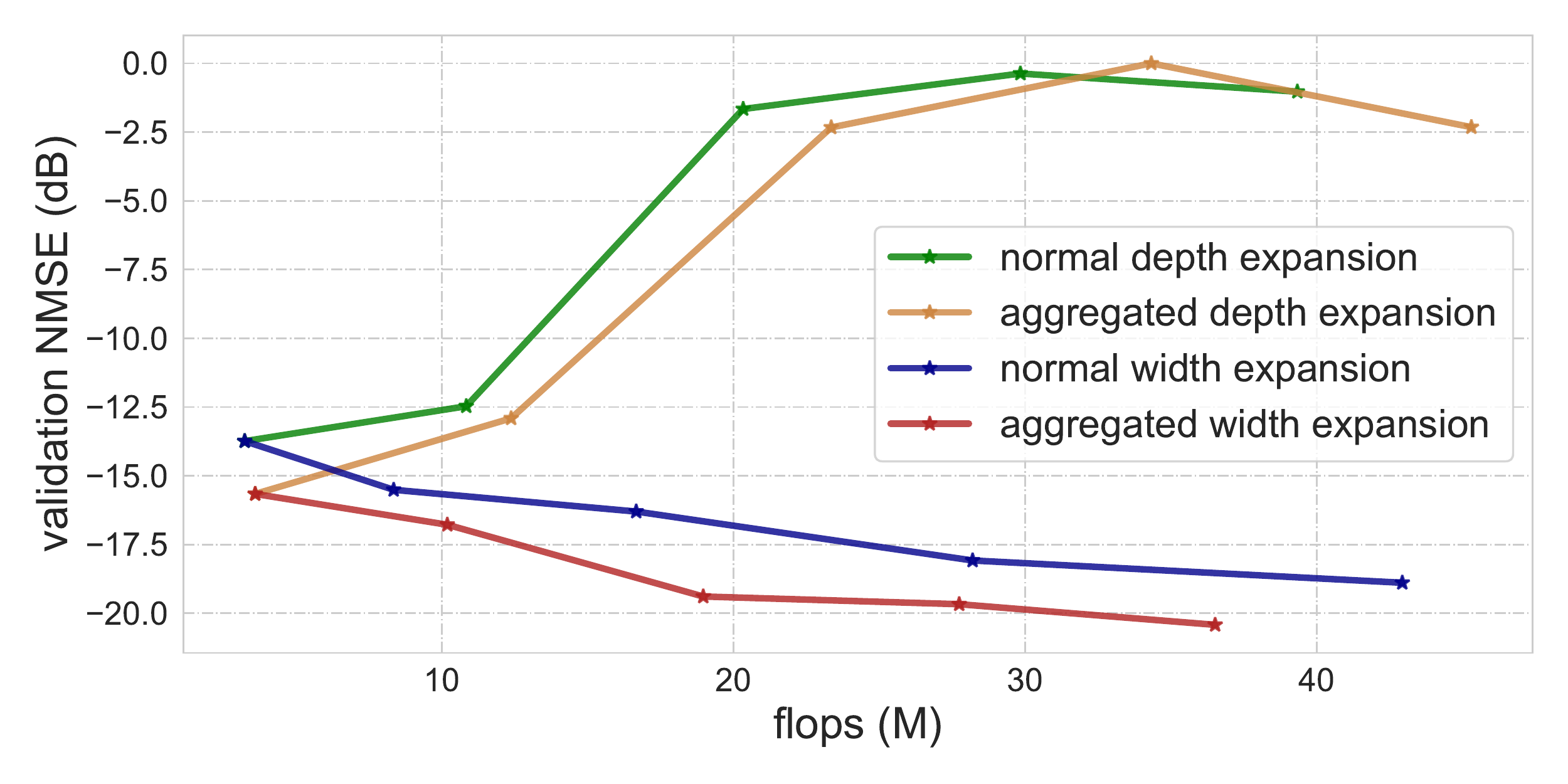}
\caption{The complexity-performance trade off for four different network expansion methods. The experiments are based on the indoor scenario and the reciprocal compression ratio is $\eta=1/8$.}
\label{Trade off}
\end{figure}

In this subsection, the effectiveness of the elastic ACRNet design will be further discussed. It is analyzed in section \ref{Section4-1} that width expansion is more suitable than depth expansion for CSI feedback task. This opinion is proved by the experiments in Fig. \ref{Trade off}, where the feedback performances of different network expansion schemes are compared.

It can be seen from Fig. \ref{Trade off} that the network gradually fails to recover the CSI matrix as it goes deeper. The performance degeneration mainly results from the loss of spatial information brought by the increasing depth. On the other hand, the network becomes more powerful with the increasing width for both normal block and aggregated block. In addition, width expansion based on the aggregated block is more efficient compared with the normal block since its validation NMSE is lower under the same complexity.

\begin{table}[!t]
\caption{NMSE (dB) and complexity comparison among expandable ACRNet variants}
\begin{center}
\makegapedcells \renewcommand\tabcolsep{10pt}
\begin{tabular}{c | r r | r r}
\Xhline{0.8pt}
\multirow{2}{*}{\textbf{Methods}} & \multicolumn{2}{c|}{\textbf{Complexity}} & \multicolumn{2}{c}{\textbf{NMSE}$^{\mathrm{a}}$} \\
& \multicolumn{1}{c}{FLOPs} & params & indoor & outdoor \\
\Xhline{0.8pt}
  ACRNet-1$\times$ & 4.64M & 2102K & -27.16 & -10.71 \\
  ACRNet-4$\times$ & 11.23M & 2109K & -28.58 & -13.13 \\
  ACRNet-8$\times$ & 20.01M & 2118K & -29.31 & -13.45 \\
  ACRNet-12$\times$ & 28.79M & 2127K & -30.28 & -13.91 \\
  ACRNet-16$\times$ & 37.58M & 2136K & \textbf{-30.81} & \textbf{-14.16} \\
\hline
\Xhline{0.8pt}
\multicolumn{5}{l}{$^{\mathrm{a}}$ $\eta$ is $1/4$ for both indoor and outdoor.} \\
\end{tabular}
\label{tab2}
\end{center}
\end{table}

\begin{figure}[!b]
\centering
\includegraphics[width=0.55\textwidth]{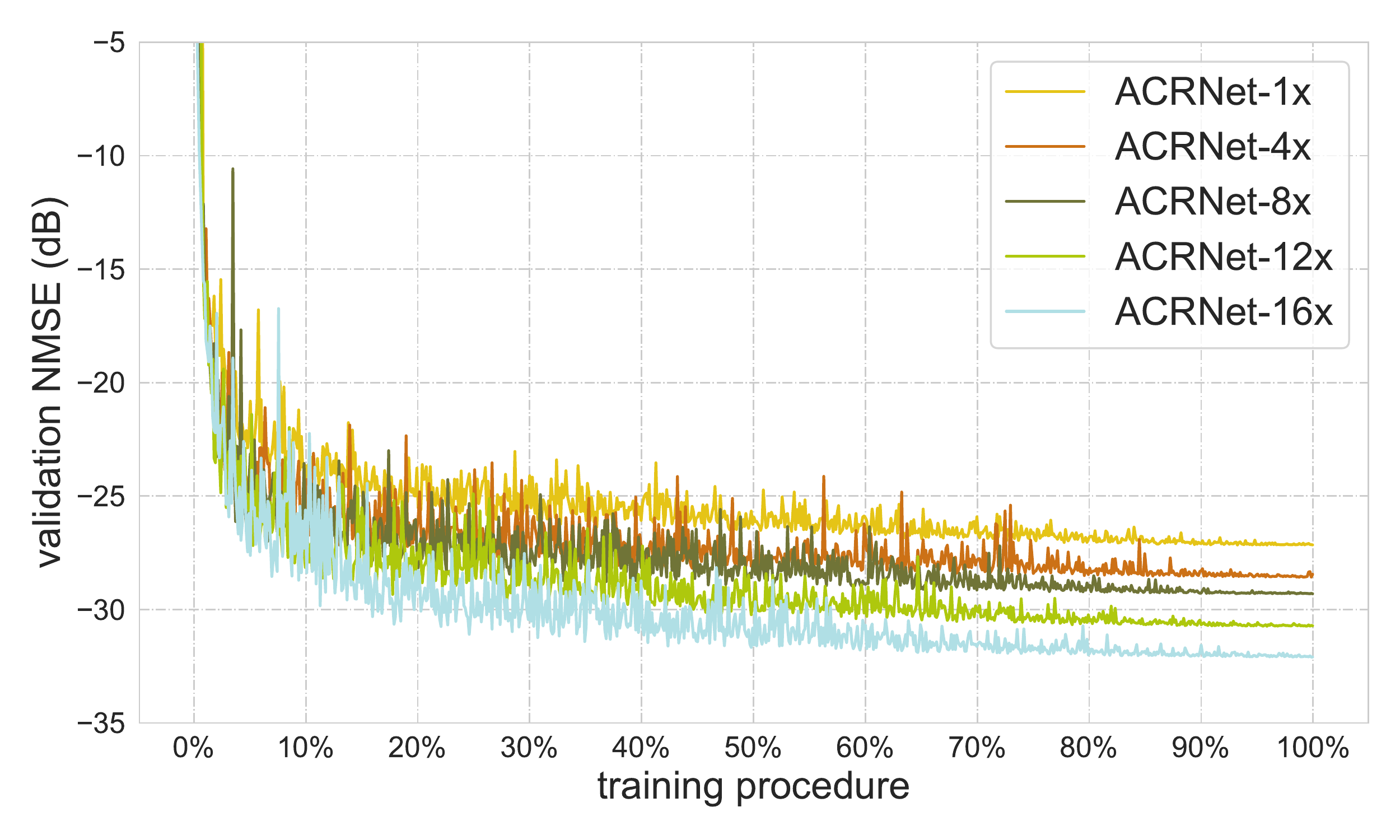}
\caption{The validation loss descending trend of CsiNet, CRNet and the proposed ACRNet variants.}
\label{Table2 NMSE}
\end{figure}

As explained in section \ref{Section4-1}, the proposed ACRNet can be easily expanded by enlarging the number of groups $M$. We test series of expanded ACRNet from ACRNet-1$\times$ ($M=4$) to ACRNet-16$\times$ ($M=64$) with $\eta=1/4$. The feedback performances at indoor and outdoor scenario are given in Table \ref{tab2}. It is clear that the larger the multiple of expansion is, the more precise the feedback will become. This proves that the proposed elastic feedback network design is quite effective.

The convergence performance of the proposed expansion scheme is shown in Fig. \ref{Table2 NMSE}. It can be seen that the network expansion does little harm to the convergence. Though the fluctuation of the NMSE curve becomes slightly larger for heavier ACRNet, the decay tendency remains stable.

It is worth mentioning that the complexity growth of the ACRNet expansion is quite reasonable. First of all, the proposed expansion scheme only enlarges the decoder at the BS. No extra burden is added to the resource sensitive encoder at the UE. Moreover, the FLOPs of the ACRNet increases linearly with the number of aggregated groups $M$. Specifically, only 2.2M extra FLOPs is required when the original ACRNet-$k\times$ is extended to ACRNet-$(k+1)\times$. The fine-grained increase in complexity makes it easy to find the proper ACRNet variant for any resource limitations, which is important for the practical deployment.

\subsection{Effectiveness of the binarization and quantization design} \label{Section5-5}

As it is mentioned in section \ref{Section4-2}, FC binarization and feature quantization are applied to reduce the memory cost and the feedback overhead. Moreover, both binarization and quantization can be combined with the elastic architecture design of ACRNet, providing practical and flexible deployment solution for CSI feedback task.

\begin{table}[!b]
\caption{NMSE(dB) and memory cost comparison of ACRNet with FC binarization}
\begin{center}
\makegapedcells \renewcommand\tabcolsep{2.3pt}
\begin{tabular}{c | c c | r r | r r | r r | r r | r r | r r}
\Xhline{0.8pt}
\multicolumn{3}{c|}{$\eta$} & \multicolumn{4}{c|}{1/4} & \multicolumn{4}{c|}{1/8} & \multicolumn{4}{c}{1/16} \\
\Xhline{0.8pt}
\multirow{2}{*}{\textbf{Methods}} &
  \multicolumn{2}{c|}{\textbf{Binary FC}} &
  \multicolumn{2}{c|}{\textbf{Params}} & \multicolumn{2}{c|}{\textbf{NMSE}} &
  \multicolumn{2}{c|}{\textbf{Params}} & \multicolumn{2}{c|}{\textbf{NMSE}} &
  \multicolumn{2}{c|}{\textbf{Params}} & \multicolumn{2}{c}{\textbf{NMSE}} \\
  & encoder & decoder &
  UE & BS & indoor & outdoor &
  UE & BS & indoor & outdoor &
  UE & BS & indoor & outdoor \\
\Xhline{0.8pt}

CsiNet \cite{wen2018deep} &  &  & 1049K & 1054K & -17.36 & -8.75 & 525K & 530K & -12.7 & -7.61 & 262K & 268K & -8.65 & -4.51 \\
BCsiNet \cite{lu2021binary} & \checkmark &  & 33K & 1054K & -17.25 & -8.35 & 17K & 530K & -12.39 & -6.26 & 8K & 268K & -8.99 & -4.17 \\
BACRNet-1$\times$ & \checkmark & & 34K & 1053K & -20.46 & -9.30 & 17K & 529K & -14.09 & -6.82 & 9K & 267K & -10.64 & -4.65 \\
BACRNet-10$\times$ & \checkmark &  & 34K & 1073K & \textbf{-23.36} & \textbf{-10.41} & 17K & 549K & \textbf{-17.47} & \textbf{-8.15} & 9K & 287K & \textbf{-12.88} & \textbf{-5.45} \\
BACRNet-1$\times$ & \checkmark & \checkmark & \textbf{34K} & \textbf{37K} & -14.20 & -7.03 & \textbf{17K} & \textbf{21K} & -11.52 & -5.52 & \textbf{9K} & \textbf{13K} & -8.83 & -2.92 \\
BACRNet-10$\times$ & \checkmark & \checkmark & \textbf{34K} & \textbf{57K} & -17.27 & -8.78 & \textbf{17K} & \textbf{41K} & -14.96 & -6.63 & \textbf{9K} & \textbf{33K} & -11.7 & -4.63 \\

\Xhline{0.8pt}

\end{tabular}
\label{table-binarization}
\end{center}
\end{table}

The performances of binarization aided ACRNet are compared with CsiNet and BCsiNet in Table \ref{table-binarization}. When the encoder FC is binarized alone, the proposed BACRNet-1$\times$ achieves better performance compared with previous state-of-the-art BCsiNet under the similar complexity. Moreover, the expanded BACRNet-10$\times$ sets a new record for CSI feedback with extremely lightweight encoder, which proves the compatibility between the binarization technique and the elastic network design. For example, NMSE for indoor scenario is reduced to -23.36 dB with only 34K encoder parameters when $\eta=1/4$. Compared with original CsiNet, over 30$\times$ memory saving is achieved at the UE and the feedback NMSE is further improved for 6 dB.

\begin{figure}[!b]
\centering
\includegraphics[width=0.55\textwidth] {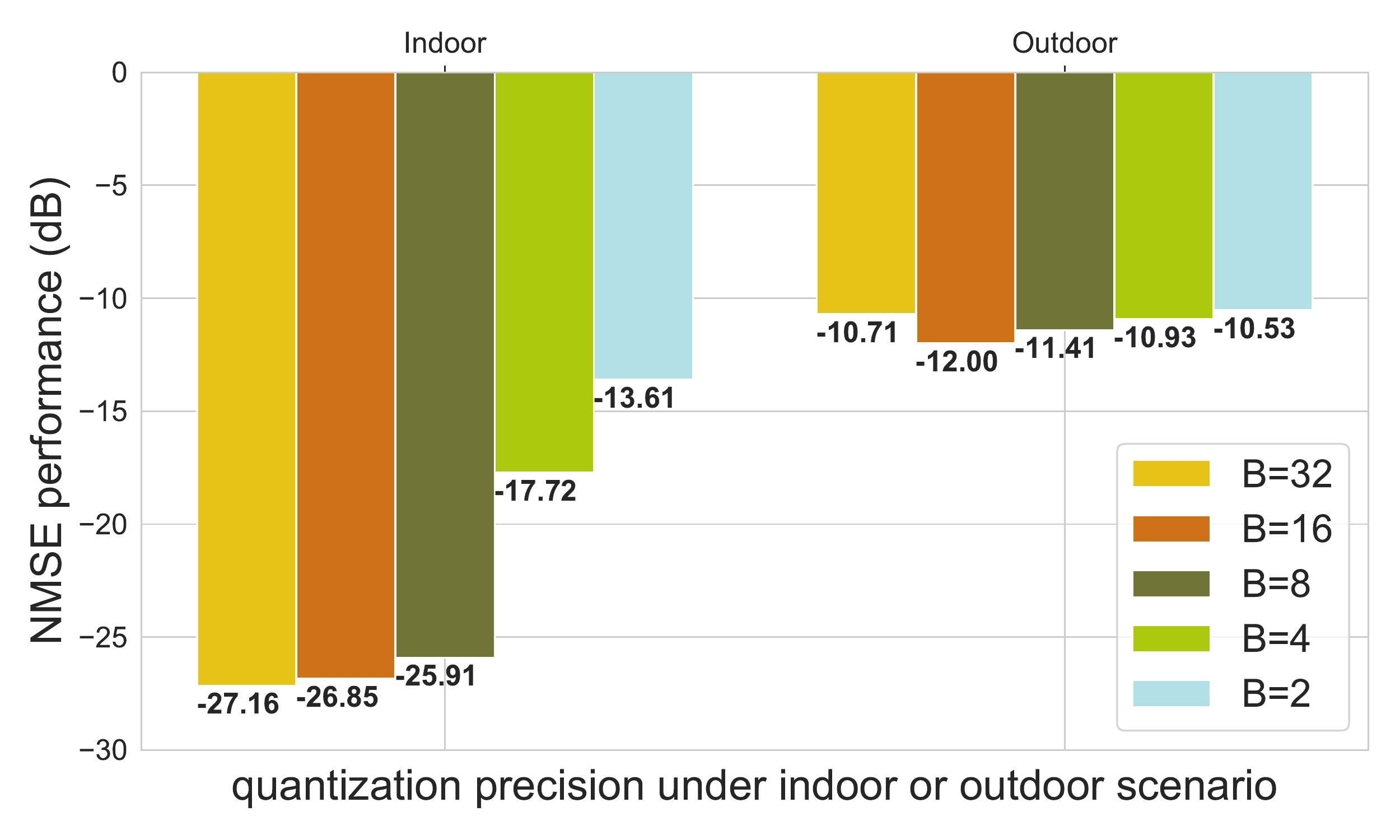}
\caption{The feedback performance of ACRNet-1$\times$ with different quantization precisions under indoor and outdoor scenario. All the performances are tested with $\eta=1/4$. Note that $B=32$ stands for the standard ACRNet-1$\times$ without quantization.}
\label{Quantization Precision}
\end{figure}

Additionally, FC binarization can be applied to ACRNet encoder and decoder at the same time. As it is shown in Table \ref{table-binarization}, the NMSE performance of BACRNet-10$\times$ is only comparable with CsiNet when both FC layers are binarized. Nevertheless, huge memory saving is achieved for both the UE and BS for the first time. For instance, around 30$\times$ and 18$\times$ memory saving are acquired respectively at the UE and BS compared with the CsiNet when $\eta=1/4$.

On the other hand, the quantization proposed in (\ref{eq-quant}) also works well with the elastic ACRNet design. The influence of quantization precision $B$ on the ACRNet-1$\times$ is given in Fig. \ref{Quantization Precision}. As we can see, the information loss with 8 bits quantization is negligible for indoor scenario when $\eta=1/4$. Moreover, the feedback performance is even better after quantization for outdoor scenario. The improvement mainly comes from the regularization effect of the quantization module.

\begin{figure}[!t]
\centering
\includegraphics[width=0.55\textwidth]{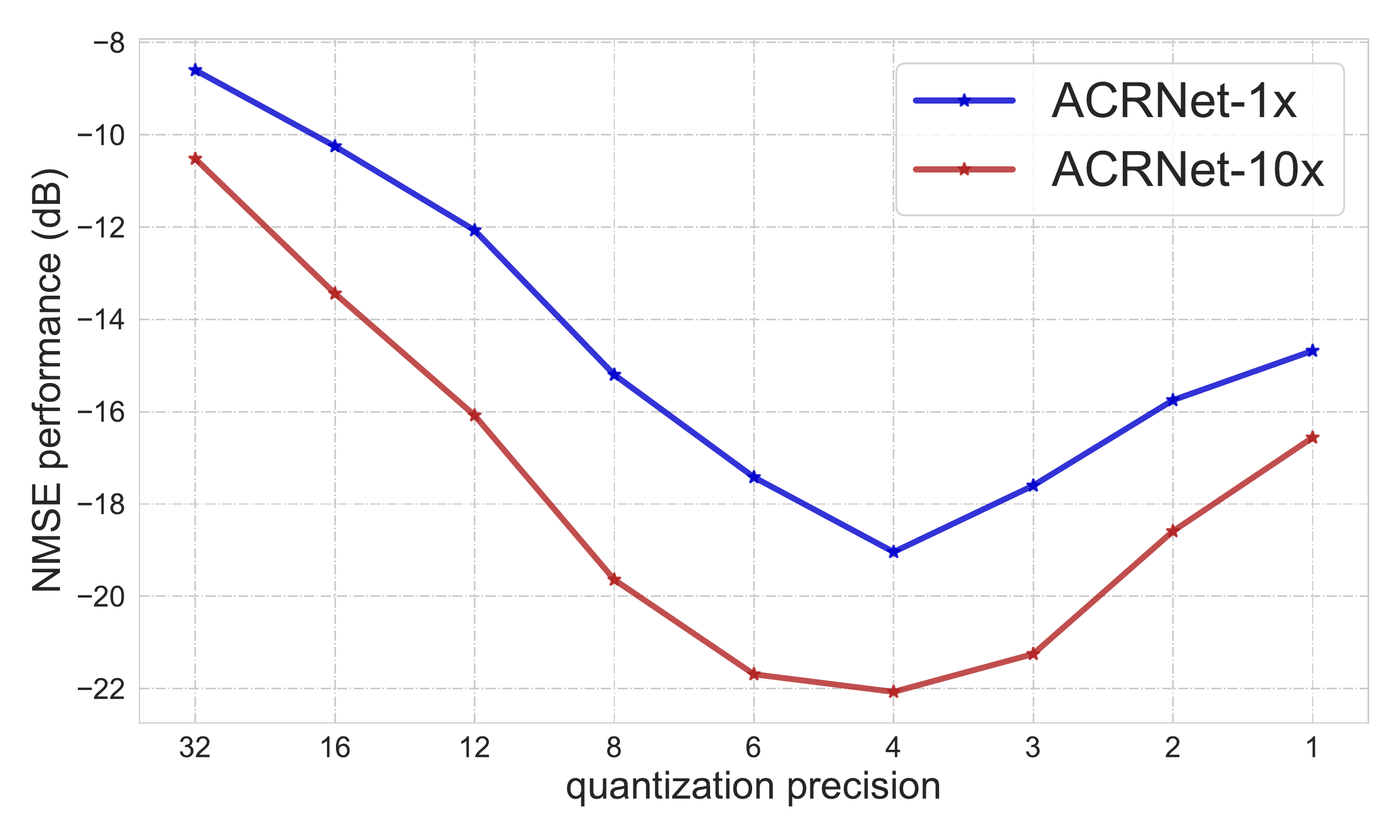}
\caption{The influence of the quantization precision $B$ on NMSE performance under the same feedback overhead. The $\eta B$ is fixed to one, therefore the bit-wise feedback overhead $N_{bf}$ is always 2048. Performances of ACRNet-1$\times$ and ACRNet-10$\times$ are studied under the indoor scenario.}
\label{Quantization Trade Off}
\end{figure}

It can be deduced from (\ref{eq-quant-trade-off}) that the feedback overhead $N_{bf}$ remains the same if the $\eta B$ is fixed. However, different combination of $\eta$ and $B$ provides different feedback performance. As it is shown in Fig. \ref{Quantization Trade Off}, best feedback performance can be achieved with $\eta=1/4$ and $B=4$ if $N_{bf}$ is limited to 2048 under the indoor scenario. Therefore, proper combination of $\eta$ and $B$ should be tested with ablation study when deploying ACRNet to a new feedback system. Note that $\eta$ smaller than $1/8$ is not recommended since the poor quality of the compressed feature $\mathbf{v}$ will become the information bottleneck.

\begin{table}[!b]
\caption{NMSE(dB) performance of quantization on BACRNet with the encoder FC binarized}
\begin{center}
\makegapedcells \renewcommand\tabcolsep{10pt}
\begin{tabular}{c c | r r | r r}
\Xhline{0.8pt}
\multirow{2}{*}{\textbf{$\bm{\eta}$}} & \multirow{2}{*}{\textbf{Bits}} & \multicolumn{2}{c|}{\textbf{Indoor NMSE}} & \multicolumn{2}{c}{\textbf{Outdoor NMSE}} \\
& & 1$\times^{\mathrm{a}}$ & 10$\times^{\mathrm{b}}$  & 1$\times$ & 10$\times$ \\
\Xhline{0.8pt}
\multirow{4}{*}{1/4}
  & 32 & -20.46 & -23.36 & -9.30 & -10.41 \\
  & 8 & -16.1 & -19.22 & -7.73 & -9.96 \\
  & 4 & -16.49 & -19.38 & -8.49 & -10.49 \\
  & 2 & -12.93 & -15.68 & -7.89 & -9.07 \\
\hline
\multirow{4}{*}{1/8}
  & 32 & -14.09 & -17.47 & -6.82 & -8.15 \\
  & 8 & -10.07 & -15.48 & -5.66 & -6.91 \\
  & 4 & -12.38 & -15.44 & -5.90 & -7.12 \\
  & 2 & -10.59 & -13.26 & -5.71 & -6.55 \\
\Xhline{0.8pt}
\multicolumn{5}{l}{$^{\mathrm{a}}$ 1$\times$ stands for BACRNet-1$\times$.} \\
\multicolumn{5}{l}{$^{\mathrm{b}}$ 10$\times$ stands for BACRNet-10$\times$.} \\
\end{tabular}
\label{table-quantization-binarization}
\end{center}
\end{table}

Finally, the compatibility of binarization, quantization and the elastic architecture design in ACRNet is studied in Table \ref{table-quantization-binarization}. Only $\eta=1/4$ and $\eta=1/8$ are considered since the performance reduction is too severe when $\eta$ is smaller than $1/8$ as it is shown in Fig. \ref{Quantization Trade Off}. Compared with the results in Fig. \ref{Quantization Precision}, the information loss brought by the quantization is increased when the encoder FC is binarized. However, huge deployment superiority can be acquired from the fusion of binarization and quantization. For example, the feedback based on the quantized BACRNet-10$\times$ can achieve -15 dB NMSE with $\eta=1/4$ and $B=2$. Combined with vanilla CsiNet with the same NMSE performance, 31$\times$ memory saving at the UE and 16 $\times$ bit-wise feedback overhead reduction are obtained at the same time.

In addition, the ACRNet-10$\times$ stably outperforms the lighter ACRNet-1$\times$ even with encoder FC binarization and vector quantization. In other words, the elastic architecture design of ACRNet still works as excepted. With the guarantee of such compatibility, the practical deployment scheme with low cost and high performance can be easily found by following the strategy proposed in Algorithm \ref{algorithm-deployment}.

\ifCLASSOPTIONcaptionsoff
  \newpage
\fi

\section{Conclusion} \label{Section6}
In this paper, a novel CSI feedback network named ACRNet was introduced for massive MIMO FDD system. The network aggregation technique and the PReLU activation were adopted and proved to be effective. The flexible deployment strategy of the proposed ACRNet was discussed as well. Effective scheme of network expansion was designed and the FC binarization was combined with the vector quantization for the first time. Experiments showed that the proposed ACRNet greatly outperformed series of previous state-of-the-art networks. Moreover, the ACRNet could be flexibly deployed with high performance, small memory cost and low feedback overhead with the given strategy.

% Reference
\bibliographystyle{IEEEtran}
\bibliography{ACRNet.bib}

% Biography
% \begin{IEEEbiography}{Yuguang ``Michael'' Fang}
% Biography text here.
% \end{IEEEbiography}

%It is not necessary to upload the biography when you submit your manuscript.

\end{document}